\documentclass[lettersize,10pt,fleqn]{article}
\usepackage{graphicx}
\usepackage{subfigure}
\usepackage{morefloats}
\usepackage{color}
\usepackage{setspace}
\usepackage{floatrow}
\usepackage{bm}
\usepackage{color}
\usepackage{float}
\usepackage{amssymb}
\usepackage{geometry}%
\usepackage{amsmath}
\usepackage{CJK} 
\usepackage[colorlinks,linkcolor=blue,citecolor=blue,bookmarks,pdfstartview=FitH]{hyperref}
\usepackage{amsthm,amsmath,amssymb}
\usepackage{mathrsfs}
\usepackage{authblk}
\usepackage{cancel}  
\usepackage{multirow}
\usepackage[bottom]{footmisc}
\usepackage{comment}
\begin{document}
\begin{sloppypar}
	
	\title{\textbf{Global destabilization of drift-tearing mode with coupling to discretized electron drift-wave instability}}
	
	\author[1,2]{J. Bao}
	\author[1,2,3,\thanks{wzhang@iphy.ac.cn}]{W. L. Zhang}
	\author[4]{Z. Lin}
	\author[5\thanks{hscai@mail.ustc.edu.cn}]{H. S. Cai}
	\author[6]{D. J. Liu}
	\author[7]{H. T. Chen} 
	\author[1,2]{C. Dong}
	\author[1,2]{J. T. Cao}
	\author[1,2,3]{D. Li}
	
	\affil[1]{\small Beijing National Laboratory for Condensed Matter Physics and CAS Key Laboratory of Soft Matter Physics, Institute of Physics, Chinese Academy of Sciences, Beijing 100190, China}
	\affil[2]{\small University of Chinese Academy of Sciences, Beijing 100049, China}
	\affil[3]{\small Songshan Lake Materials Laboratory, Dongguan, Guangdong 523808, China}
	\affil[4]{\small University of California, Irvine, California, 92697, USA}
	\affil[5]{\small CAS Key Laboratory of Geospace Environment, School of Nuclear Sciences and echnology, University of Science and Technology of China, Hefei 230026, China}
	\affil[6]{\small College of Physical Science and Technology, Sichuan University, Chengdu 610064, China}
	\affil[7]{\small Fusion Simulation Center, Peking University, Beijing 100871, China}
	
	\date{}
	
	\maketitle
	\begin{abstract}
		The global linear behaviors of $m/n=2/1$ drift-tearing mode (DTM) in the collisional regime are investigated based on a concisely resistive drift-MHD model. Besides DTM, extra normal modes including electron drift-wave (EDW) and shear Alfv\'en wave (SAW) are coupled together and destabilized in different parameter regimes by considering resistivity in this system. The eigenvalue problem (EVP) approach is applied for solving the eigenstate spectra with the distribution of all unstable solutions on $\omega_r$-$\gamma$ complex plane. It is found that in the small electron diamagnetic drift (EDD) frequency regime $|\omega_{*n,e}|\leq\gamma_c$ ($\gamma_c$ is the classical resistive-tearing mode (RTM) growth rate), DTM growth rate agrees well with local theory that is reduced with increasing EDD frequency. However, when EDD frequency exceeds a critical threshold $|\omega_{*n,e}|> |\omega_{*n,e}^{crit}|\sim 2\gamma_c$, the strongly linear coupling between DTM and other discretized EDW instabilities happens so that the free energies from current and pressure channels can be released together and thus enhance the DTM instability, of which growth rate increases with increasing EDD frequency and deviates from local theory qualitatively. Correspondingly, a cross-scale mode structure forms with mixed polarization, namely, $\delta\phi$ perturbation is dominated by electrostatic polarized short-wavelength oscillation as EDW instability character, and $\delta A_{||}$ perturbation remains typical tearing mode solution of Alfv\'enic polarized macroscopic structure. Within $|\omega_{*n,e}|>|\omega_{*n,e}^{crit}|$, the additional ion diamagnetic drift (IDD) causes $\delta\phi$ oscillating structure to shift towards small density gradient domain, which cancels the extra drive from ion channel and thus DTM growth rate is insensitive to IDD frequency. Compared to EDD effects, the IDD effect alone with zero-$\omega_{*n,e}$ only leads to the stabilization of RTM that shows agreements between global simulation and local theory, which is no longer the condition for DTM regime. These results are useful for clarifying the DTM global properties with underlying physics mechanisms, which occurs in the regime of $|\omega_{*n,e}|\gg\gamma_c$ that is relevant to nowadays tokamak discharges with hot plasmas. 
	\end{abstract}
	
	\newpage
	
	\section{Introduction}
	
	Tearing mode is a common current-driven magnetohydrodynamics (MHD) instability in space and laboratory plasmas that attracts community attention for decades, of which process can be either collisional or collisionless depending on the relative importance of physics terms in Ohm's law. The original theory is established for resistive tearing mode (RTM) based on asymptotic matching method by Furth et al in Ref. \cite{Furth1963}, which utilizes constant-$\psi$ approximation and predicts the growth rate scaling with resistivity as $\gamma\sim \eta^{3/5}$. In high temperature fusion plasmas, kinetic effects beyond MHD, such as electron/ion diamagnetic drift (EDD/IDD), are shown to be important that can change RTM characters into drift-tearing mode (DTM) regime, and various DTM theories with different level physics models are proposed and generally indicate that finite EDD frequency (i.e, $\omega_{*n,e}$) not only reduces the mode growth rate in scaling of $\gamma\sim |\omega_{*n,e}|^{-2/3}$ but also induces a real frequency close to $\omega_{*n,e}$ \cite{Coppi1964,Coppi1965, Rutherford1971, Hazeltine1975, Drake1977}. Besides the DTM destabilizing mechanisms existing in slab geometry, it has been shown analytically by Chen et al that the free energy of trapped electrons can also drive DTM in the collisional regime based on the drift-kinetic model in toroidal geometry \cite{Chen1977}. 
	
	Besides the efforts in theory, various numerical studies have been performed to investigate DTM linear stability \cite{Biskamp1978, Grasso2001, Grasso2002, Shi2019, Shi2020} and nonlinear evolution \cite{Monticello1980, Porcelli2001, Ottaviani2004, Fitzpatrick2008, Yu2010, Ye2019}, which further reveal the complex roles of EDD and IDD on tearing mode physics beyond theory. For example, the dependence of DTM growth rate on $\omega_{*n,e}$ from global simulations exhibits strong deviation from theoretical scaling of $\gamma\sim|\omega_{*n,e}|^{-2/3}$, and a nonlocalized mode structure of drift type appears being different from typical tearing mode structure \cite{Biskamp1978, Grasso2001}. Therefore, the complexity of DTM problem not only attributes to the well-known plasma nonlinearities, but also comes from the global effects that even alter the linear stability and mode structure qualitatively. Although traditional DTM theories \cite{Coppi1964,Coppi1965,Rutherford1971,Hazeltine1975,Drake1977} have already considered EDD and IDD effects in deriving the dispersion relation, the co-existing electron drift wave (EDW) can also become unstable in presence of pressure gradient drive and plasma resistivity \cite{Horton1999}, and the mutual coupling between DTM and EDW instability in different scales can be complicated and is not considered in former theory. In general, the dispersion relation and mode structure are connected to each other and reflect the overall stability of eigenstate. The scope of this work is to numerically investigate the global behaviors of linear DTM including both dispersion relation and mode structure, which will also be compared with local theory obtained from the identical physics model.
	
	Most DTM numerical studies utilize the initial value simulations that self-consistently time advance the dynamic equations incorporating the linear and nonlinear terms on the same footing, which have advantages on nonlinear problems, while only the most unstable solution can be obtained. For comparison, the eigenvalue problem (EVP) approach is able to compute all unstable solutions and give eigenvalue distribution on $\omega_r-\gamma$ complex plane, which is more efficient for linear stability analysis. The MAS eigenvalue code has been developed for studying various plasma instabilities in general geometry, which consists of multi-level physics models up to five-field Landau-fluid model for bulk plasmas \cite{Bao2023} and the drift-kinetic model for energetic-electrons \cite{Bao2024,Bao2023b}. In this work, the drift-MHD model in MAS framework is adopted for DTM study so that the technical complications can be minimized for demonstrating global nature exclusively. Specifically, the theoretical dispersion relations of both DTM and normal modes of EDW/shear Alfven wave (SAW) co-existing in the system are derived, and corresponding growth rate and real frequency variations with plasma parameters are analyzed. Global MAS simulations are carried out to obtain the unstable solution spectra based on identical model equations, and the characters of global DTM with coupling to EDW instability are studied in detail, including the non-monotonic varied dispersion relation with both stabilizing and destabilizing regimes on $|\omega_{*n,e}|$ and cross-scale mixed mode structure with both electrostatic and Alfv\'enic polarizations, as well as the role of IDD effect in the coupling process.
	
	This work is organized as follows. The drift-MHD physics model is introduced in section 2. The theoretical dispersion relations of DTM and normal modes of EDW and SAW are derived and solved in parameter space in section 3. MAS simulation results of global DTM dispersion relation and mode structure are shown in section 4. The conclusion and discussion are given in section 5.

	\section{Drift-MHD physics model}\label{drift_MHD}
 The ion and electron diamagnetic-drift effects are found to be important for plasma waves and instabilities adapting to drift-ordering \cite{Hazeltine_book, Schnack2006}
	\begin{flalign}\label{ordering}
	\begin{split}
	 \mathbf{V_\perp}\sim\mathbf{V_E}\sim\mathbf{V_{i}^*}\sim\mathbf{V_{e}^*}
	\end{split},
\end{flalign}
 where $\mathbf{V}_\perp$ represents the perpendicular plasma flow, $\mathbf{V_E}=c\mathbf{E}\times\mathbf{B}/{B^2}$ is the $E\times B$ drift, $\mathbf{V_i^*}=c\mathbf{B}\times\nabla P_i/\left(Z_iN_iB^2\right)$ and  $\mathbf{V_e^*}=c\mathbf{B}\times\nabla P_e/\left(q_eN_eB^2\right)$ denote the ion and electron diamagnetic-drifts respectively. To delineate the global nature of $m/n=2/1$ DTM stability in a concise manner, a reduced-resistive-drift-MHD model in the lowest order of $O\left(\omega_{*p,i(e)}/\omega\right)$ is used, of which model equations are expressed as follows

	\begin{flalign}\label{vor3}
		\begin{split}
			&\frac{\partial }{\partial t}\frac{c}{V_A^2}\nabla_\perp^2\delta\phi
			\underbrace{+
				i\omega_{*p,i}\frac{c}{V_A^2}\nabla_\perp^2\delta\phi}_{\{Ion\ dia-drift\}}
			+\mathbf{B_0}\cdot\nabla\left(\frac{1}{B_0}\nabla_\perp^2\delta A_{||}\right) 
			-\frac{4\pi}{c}\boldsymbol{\delta B}\cdot\nabla\left(\frac{J_{||0}}{B_0}\right)
			=0,
		\end{split}
	\end{flalign}
	\begin{flalign}\label{ohm3}
		\begin{split}
			\frac{\partial \delta A_{||}}{\partial t} =
			& -c\mathbf{b_0}\cdot\nabla\delta \phi 
			\underbrace{+ \frac{cT_{e0}}{en_{e0}}\mathbf{b_0}\cdot\nabla\delta n_e
				+ \frac{cT_{e0}}{en_{e0}B_0}\boldsymbol{\delta B}\cdot\nabla n_{e0}}_{\{Electron\ dia-drift\}}
			\underbrace{+ \frac{c^2}{4\pi}\eta_{||}\nabla_\perp^2\delta A_{||}}_{\{Resistivity\}}
		\end{split},
	\end{flalign}
	\begin{flalign}\label{dne3}
		\begin{split}
			\frac{\partial \delta n_e}{\partial t} 
			-i\frac{q_en_{e0}}{T_{e0}}\omega_{*n,e}\delta\phi
			=0
		\end{split},
	\end{flalign}
	where $e$ and $c$ denote the elementary charge and light speed respectively. $\delta\phi$ is the electrostatic potential, $\delta A_{||}$ is the parallel vector potential, $\mathbf{B_0}$ is the equilibrium magnetic field and $\mathbf{\delta B} = \nabla\delta A_{||}\times \mathbf{b_0}$ is the perturbed magnetic field. $\omega_{*p,i} =\omega_{*n,i}  + \omega_{*T,i} $ is the ion diamagnetic frequency, $\omega_{*n,i} = -i\frac{cT_{i0}}{Z_iB_0}\mathbf{b_0}\times\frac{\nabla n_{i0}}{n_{i0}}\cdot\nabla$ and $\omega_{*T,i} = -i\frac{c}{Z_iB_0}\mathbf{b_0}\times\nabla T_{i0}\cdot\nabla$, $T_{i0}$ and $n_{i0}$ are the ion equilibrium temperature and density respectively. $\omega_{*n,e} = -i\frac{cT_{e0}}{q_eB_0}\mathbf{b_0}\times\frac{\nabla n_{e0}}{n_{e0}}\cdot\nabla$ is the electron diamagnetic drift frequency due to density gradient, and uniform electron temperature is considered for simplicity, $T_{e0}$ and $n_{e0}$ are the electron equilibrium temperature and density respectively, and $\delta n_{e}$ is the electron perturbed density. $V_A = B_0/\sqrt{4\pi n_{i0}m_i}$ is the ion Alfv\'en speed. $J_{||0} = \frac{c}{4\pi}\mathbf{b_0}\cdot\nabla\times\mathbf{B_0}$ is the parallel equilibrium current density, $\eta_{||} = 0.51\frac{m_e\nu_{ei}}{n_{e0}e^2}$ is the resistivity coefficient and $\nu_{ei}$ is the electron-ion collision frequency. This model describes the drift-MHD dynamics for plasmas in a low-$\beta$ and high-aspect-ratio tokamak where the interchange drive generated by the coupling between pressure gradient and magnetic field curvature is dropped in Eq. \eqref{vor3} as higher order term for drift-tearing mode, while the necessary ion diamagnetic-drift term in Eq. \eqref{vor3} and parallel gradient of electron pressure in Eq. \eqref{ohm3} are kept and consistent to Hazeltine's four-field model in Ref. \cite{Hazeltine85}. Moreover, the isothermal condition is applied for electron species in Eq. \eqref{ohm3} with $\nabla_{||} T_e = \mathbf{b_0}\cdot\nabla\delta T_e + \frac{1}{B_0}\mathbf{\delta B}\cdot\nabla T_{e0} = 0$, which is appropriate for the Alfv\'en-time scale waves \cite{Snyder2001} and brings the simplification of $\nabla_{||} P_e=\mathbf{b_0}\cdot\nabla\delta P_e + \frac{1}{B_0}\mathbf{\delta B}\cdot\nabla P_{e0} =T_{e0}\mathbf{b_0}\cdot\nabla\delta n_e + \frac{T_{e0}}{B_0}\mathbf{\delta B}\cdot\nabla n_{e0}$. It is also worthwhile mentioning that this drift-MHD model described by Eqs. \eqref{vor3}-\eqref{dne3} covers the well-known reduced-MHD model in Ref. \cite{Strauss76} in the limits of $\omega_{*n,e}\to 0$, $\omega_{*p,i}\to 0$ (i.e., zero-$\beta$) and $\eta_{||}\to 0$. 
	
	Inspired by recent linear stability analysis of collisionless double tearing mode \cite{Wang2011}, the eigenvalue problem (EVP) approach and drift-MHD model in MAS multi-level physics framework \cite{Bao2023} are applied for studying DTM physics in this work, which takes both the physics advantage of obtaining all unstable solutions and computational advantage of high efficiency in a broad range of parameter regime. In particular, the EVP approach can show the eigenvalue distribution on $\omega_r-\gamma$ complex plane and reveal the correlation between different branch modes intuitively. The detailed numerical scheme and the implementation of drift-MHD model in MAS code are introduced in Appendix \ref{app_A1}.

   \section{Theoretical dispersion relation}\label{theory}
   Though Eqs. \eqref{vor3}-\eqref{dne3} consists of the minimal model for describing DTM physics \cite{Grasso2001}, there still exist other normal modes in the system. In this section, we derive the dispersion relation and carry out stability analysis for each branch mode based on drift-MHD model, which aims to provide theoretical references for comparing with and understanding global simulation results in section \ref{simulation}.
   
   \subsection{Discretized eigenmode: drift-tearing mode}\label{theory_DTM}
To confirm that DTM physics is contained in our physics model, we first derive its linear dispersion relation in a cylinder geometry according to Eqs. \eqref{vor3}-\eqref{dne3}, and the perturbation is represented as $U\left(r,\theta,z,t\right)=\frac{1}{2} \sum_{m,n}U_{m,n}exp\left(-im\theta+inz/R_0-i\omega t\right) + c.c.$, where $m$ and $n$ are the poloidal and axial harmonic numbers, and $c.c.$ represents the complex conjugate. Applying the Fourier transform $\partial_t\to -i\omega$ and $\mathbf{b_0}\cdot\nabla = ik_{||}$, and defining the normalizations 
\begin{flalign}\label{DTM_normalization}
	\begin{split}
		\left(\widetilde{\delta\phi}, \widetilde{\delta A}_{||}, \widetilde{t}, \widetilde{\nabla}, \widetilde{J}_{||0}, \widetilde{\eta}_{||}\right)
		=\left(\frac{c\delta\phi}{B_0V_AR_0}, \frac{\delta A_{||}}{B_0R_0}, \frac{V_A}{R_0}t, R_0\nabla, \frac{4\pi R_0}{cB_0}J_{||0}, \frac{c^2\eta_{||}}{4\pi V_AR_0}\right)
	\end{split},
\end{flalign}
Eqs. \eqref{vor3}-\eqref{dne3} can be expressed as coupled equations for $\widetilde{\delta\phi}$ and $\widetilde{\delta A}_{||}$
   	\begin{flalign}\label{vor_norm}
   	\begin{split}
   		&-i\left(\widetilde{\omega}- \widetilde{\omega}_{*p,i}\right)\widetilde{\nabla}_\perp^2\widetilde{\delta\phi}
   		+i\widetilde{k}_{||}\widetilde{\nabla}_\perp^2\widetilde{\delta A}_{||}
   		=-i\widetilde{k}_\theta\widetilde{\delta A}_{||}\frac{d\widetilde{J}_{||0}}{d\widetilde{r}}
   	\end{split},
   \end{flalign}
   	\begin{flalign}\label{ohm_norm}
   	\begin{split}
   		-i\left(\widetilde{\omega}-\widetilde{\omega}_{*n,e}\right)\widetilde{\delta A}_{||} =
   		& -\left(1-\frac{\widetilde{\omega}_{*n,e}}{\widetilde{\omega}}\right)i\widetilde{k}_{||}\widetilde{\delta \phi} 
   	+ \widetilde{\eta}_{||}\widetilde{\nabla}_\perp^2\widetilde{\delta A}_{||}
   	\end{split},
   \end{flalign}
   where $\widetilde{k}_\theta = m/\widetilde{r}$ is the poloidal wave vector. For simplicity of notation, we omit the normalized symbol in the following derivation. 
   
Following the well-established RTM theory \cite{Furth1963, Furth1973}, the asymptotic matching method is applied to solve the physics equations separately in the inner and outer regions regarding to the resistive layer, and then connect to each other through the boundary condition. Besides the applications on bulk plasma related issues, recently this method has been extended to study the runaway electron effect on resistive-MHD mode \cite{Liuchang2020}. In the inner region close to the mode rational surface with $k_{||} = 0$, we induce a new radial coordinate $x=r-r_s$ with $r_s$ being the minor radius at rational surface, and $k_{||} \approx nsx/r$ for $|x|\ll r_s$, where $s=\left(r/q\right)\left(dq/dr\right)$. Considering the approximation of $\nabla_\perp^2\approx d^2/dr^2$ for perturbed field with fast radial variation, and applying the constant-$\psi$ approximation of $\delta A_{||}(x) \approx \delta A_{||}(0)$ in the inner region with $k_{||}\sim 0$,  Eqs. \eqref{vor_norm} and \eqref{ohm_norm} reduce to
   	\begin{flalign}\label{inner_region}
   	\begin{split}
   	    -i\eta_{||}\frac{\omega-\omega_{*p,i}}{\omega-\omega_{*n,e}}\omega\frac{d^2\delta\phi\left(x\right)}{dx^2} 
   	    -\left(\frac{nsx}{r}\right)^2\delta\phi\left(x\right)
   	    + \frac{nsx}{r}\omega\delta A_{||}(0)=0
   	\end{split}.
   \end{flalign} 
   Introduce the transformations
   	\begin{flalign}\label{DTM_trans_phi}
   	\begin{split}
   	   \delta\phi\left(x\right)
   	   =-\left[\frac{1}{\left(-i\omega\eta_{||}\right)}\frac{\omega-\omega_{*n,e}}{\omega-\omega_{*p,i}}\left(\frac{r}{ns}\right)^2\right]^{1/4}\omega\delta A_{||}(0)\chi\left(z\right)
   	\end{split}
   \end{flalign} 
   and
   	\begin{flalign}\label{DTM_trans_x}
   	\begin{split}
   	x = \frac{r}{ns}\left[-i\omega\eta_{||}\frac{\omega-\omega_{*p,i}}{\omega-\omega_{*n,e}}\left(\frac{ns}{r}\right)^2\right]^{1/4}z
   	\end{split},
   \end{flalign} 
   and substitute into Eq. \eqref{inner_region}, we have
   	\begin{flalign}\label{}
   	\begin{split}
   		\frac{d^2\chi\left(z\right)}{dz^2} - z^2\chi\left(z\right) = z
   	\end{split},
   \end{flalign} 
  of which solution is $\chi\left(z\right) = -\left(z/2\right)\int_{0}^{1}d\mu exp\left(-z^2\mu/2\right)\left(1-\mu^2\right)^{-1/4}$, and the integral required for the calculation of matching condition can be analytically derived as $\int_{-\infty}^{+\infty}\frac{d^2\chi\left(z\right)}{dz^2}\frac{dz}{z} = 2\pi\Gamma\left(3/4\right)/\Gamma\left(1/4\right)$ \cite{White1983}. Since $dJ_{||0}/dr$ term on the RHS of Eq. \eqref{vor_norm} is much smaller than other terms, the relation between second order derivative of electromagnetic fields with respect to $x$ in the inner region can be obtained as
  	\begin{flalign}\label{d2Adx2}
  	\begin{split}
  	\frac{d^2\delta A_{||}\left(x\right)}{dx^2} = \frac{\omega-\omega_{*p,i}}{k_{||}}\frac{d^2\delta\phi\left(x\right)}{dx^2}
  	\end{split}.
  \end{flalign}
 From Eqs. \eqref{DTM_trans_phi}, \eqref{DTM_trans_x} and \eqref{d2Adx2}, one can match the inner region solution to the outer region one as
   	\begin{flalign}\label{DTM_delta}
   	\begin{split}
   	 \Delta'
   	 &=\frac{1}{\delta A_{||}(0)}\int_{-\infty}^{+\infty}\frac{d^2\delta A_{||}\left(x\right)}{dx^2} dx\\
   	 &=\left(\frac{r}{ns}\right)\frac{\omega-\omega_{*p,i}}{\delta A_{||}(0)}\int_{-\infty}^{+\infty}\frac{d^2\delta \phi\left(x\right)}{dx^2} \frac{dx}{x}\\
   	 &=-\left(\frac{ns}{r}\right)^{-1/2}\eta_{||}^{-3/4}\omega\left(\omega-\omega_{*p,i}\right)\left[i\frac{\left(\omega-\omega_{*n,e}\right)}{\omega\left(\omega-\omega_{*p,i}\right)}\right]^{3/4}2\pi\frac{\Gamma\left(3/4\right)}{\Gamma\left(1/4\right)}
   	\end{split},
   \end{flalign}
where $\Delta'$ is the stability criterion of tearing mode that is determined by the outer region solution. From Eq. \eqref{DTM_delta}, the local dispersion relation of DTM can be readily derived as
    \begin{flalign}\label{DR_DTM}
   	\begin{split}
   	 \omega\left(\omega - \omega_{*p,i}\right)\left(\omega-\omega_{*n,e}\right)^3 = i\gamma_c^5
   	\end{split},
   \end{flalign}
   where 
       \begin{flalign}\label{DR_RTM}
   	\begin{split}
   		\gamma_c = \eta_{||}^{3/5}\left(\frac{ns}{r}\right)^{2/5}\left[\Delta'\frac{\Gamma\left(1/4\right)}{2\pi\Gamma\left(3/4\right)}\right]^{4/5}
   	\end{split}
   \end{flalign}
 is the classical RTM growth rate. Note that the physics quantities in Eq. \eqref{DR_RTM} are normalized ones according to Eq. \eqref{DTM_normalization}.
 
 From Eqs. \eqref{DR_DTM} and \eqref{DR_RTM}, it is seen that Eqs. \eqref{vor3}-\eqref{dne3} form the most concise physics model for DTM study by only keeping the finite-$\beta$ effects of necessary EDD/IDD frequency on top of resistive reduced-MHD part while isolating the magnetic curvature coupling (i.e., no GGJ-effect \cite{Glasser1975}), which is helpful for revealing the global nature of DTM stability exclusively. To compare with the well-known drift-kinetic model result, Eq. (84a) in Ref. \cite{Drake1977} can reduce to Eq. \eqref{DR_DTM} by taking approximations of constant electron temperature and real parallel conductivity that are consistent with this work model assumption. It should be pointed out that the charge sign has been incorporated in the diamagnetic-drift frequency in our work, thus the minus sign appears in front of $\omega_{*p,i}$ in Eq. \eqref{DR_DTM} different from the plus sign in Eq. (84a) of Ref. \cite{Drake1977}. 
 
\begin{figure}[H]
   	\center
   	\includegraphics[width=0.8\textwidth]{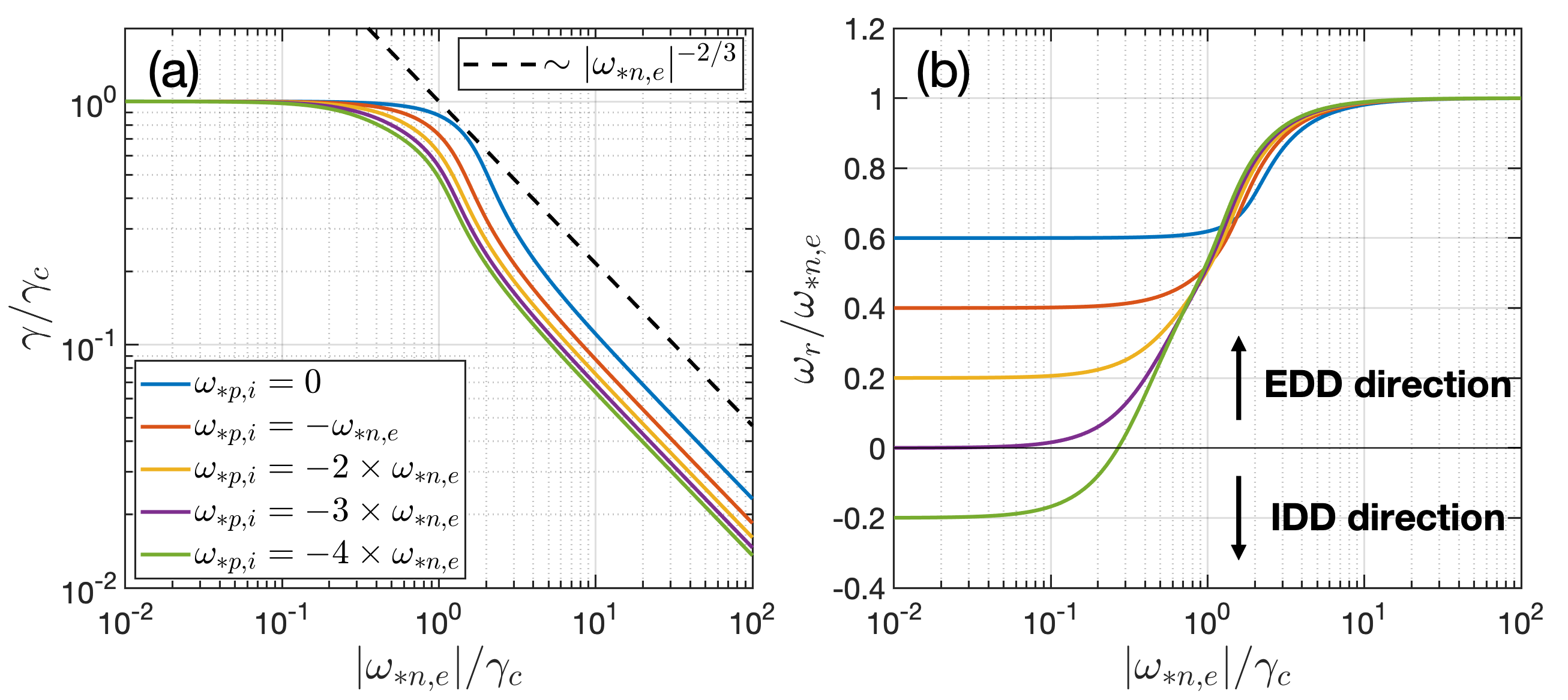}
   	\caption{DTM dispersion relation based on Eq. \eqref{DR_DTM}. The dependences of normalized (a) growth rate $\gamma/\gamma_c$ and (b) real frequency $\omega_r/\omega_{*n,e}$ on normalized EDD frequency amplitude $|\omega_{*n,e}|/\gamma_c$.}
   	\label{DTM_theory_fig}	
   \end{figure}

 Then we solve Eq. \eqref{DR_DTM} and analyze local DTM stability as shown in figure \ref{DTM_theory_fig}. It is found that (i) the DTM growth rate decreases with increasing $|\omega_{*n,e}|$, which has the scaling of $\gamma\sim |\omega_{*n,e}|^{-2/3}$ in the regime of $|\omega_{*n,e}|/\gamma_c\gg 1$. (ii) The DTM real frequency slowly varies between $\omega_r = \left(3\omega_{*n,e}+\omega_{*p,i}\right)/5$ in the lower limit of $|\omega_{*n,e}|/\gamma_c\ll 1$ and $\omega_r = \omega_{*n,e}$ in the upper limit of $|\omega_{*n,e}|/\gamma_c\gg 1$. (iii) The finite $\omega_{*p,i}$ effect shifts the growth rate curve towards the smaller $|\omega_{*n,e}|/\gamma_c$ regime and thus induces an earlier stabilization. (iv) For the moderate values of $|\omega_{*n,e}|/\gamma_c$ and below, the finite $\omega_{*p,i}$ effect leads to the decrease of DTM real frequency in EDD direction and further transits to IDD direction at a critical value $\omega_{*p,i}=-3\omega_{*n,e}$. Interestingly, as $|\omega_{*n,e}|/\gamma_c$ increases, the DTM real frequency eventually reaches to the limiting result of $\omega_r=\omega_{*n,e}$ in the regime of $|\omega_{*n,e}|/\gamma_c\gg 1$, which is independent of $\omega_{*p,i}$. 

   \subsection{Normal modes: electron drift wave and shear Alfv\'en wave}\label{theory_EDW_SAW}
    Besides DTM solution, two normal modes, i.e., electron drift wave (EDW) and shear Alfv\'en wave (SAW), are also included in drift-MHD model in section \ref{drift_MHD}, which couple with each other and become unstable in presence of finite plasma resistivity $\eta_{||}>0$. Here, we derive the theoretical dispersion relation of coupled EDW and SAW in uniform plasma and equilibrium magnetic field. Dropping the equilibrium current term (i.e., $J_{||0}$ term) and applying Fourier transformations $\partial_t\to -i\omega$ and $\nabla\to i\mathbf{k}$, one can derive the normal mode dispersion relation based on Eqs. \eqref{vor3}-\eqref{dne3}
  \begin{flalign}\label{DR_DAW}
	\begin{split}
		\left(\omega-\omega_{*n,e}\right)\left[ 1-\frac{k_{||}^2V_A^2}{\omega\left(\omega-\omega_{*p,i}\right)}\right] = -i\frac{c^2}{4\pi}\eta_{||}k_{\perp}^2
	\end{split}.
\end{flalign}
From Eq. \eqref{DR_DAW}, it is clearly seen that plasma resistivity not only leads to the coupling between EDW branch (i.e., $\omega=\omega_{*n,e}$) and two SAW branches (i.e., $\omega\left(\omega-\omega_{*p,i}\right)=k_{||}^2V_A^2$), but also give rise to the destabilization of these normal modes in certain parameter regime. For example, EDW and SAW couple through resistivity with additional magnetic field curvature effect that become unstable, are termed as drift Alfv\'en wave instability being responsible for the weakly coherent mode at the edge of I-mode plasmas found by BOUT++ simulations \cite{LiuZixi2016, Lang2022}. For analysis convenience, Eq. \eqref{DR_DAW} is casted into the dimensionless form as
\begin{flalign}\label{DR_DAW2}
	\begin{split}
		\left(\hat{\omega}-\hat{\omega}_{*n,e}\right)\left[ 1-\frac{1}{\hat{\omega}\left(\hat{\omega}-\hat{\omega}_{*p,i}\right)}\right] =  -i\frac{1}{S}\frac{k_{\perp}^2}{k_{||}^2}
	\end{split},
\end{flalign}
where the frequencies are normalized with $\hat{\omega} = \omega/(k_{||}V_A)$, $\hat{\omega}_{*n,e} = \omega_{*n,e}/(k_{||}V_A)$ and $\hat{\omega}_{*p,i} = \omega_{*p,i}/(k_{||}V_A)$, and $S = 4\pi V_A/(c^2\eta_{||}k_{||})$ is the Lundquist number. Next, Eq. \eqref{DR_DAW2} is solved numerically to check the key parameter influences on SAW/EDW normal mode stability.

\begin{figure}[H]
	\center
	\includegraphics[width=0.8\textwidth]{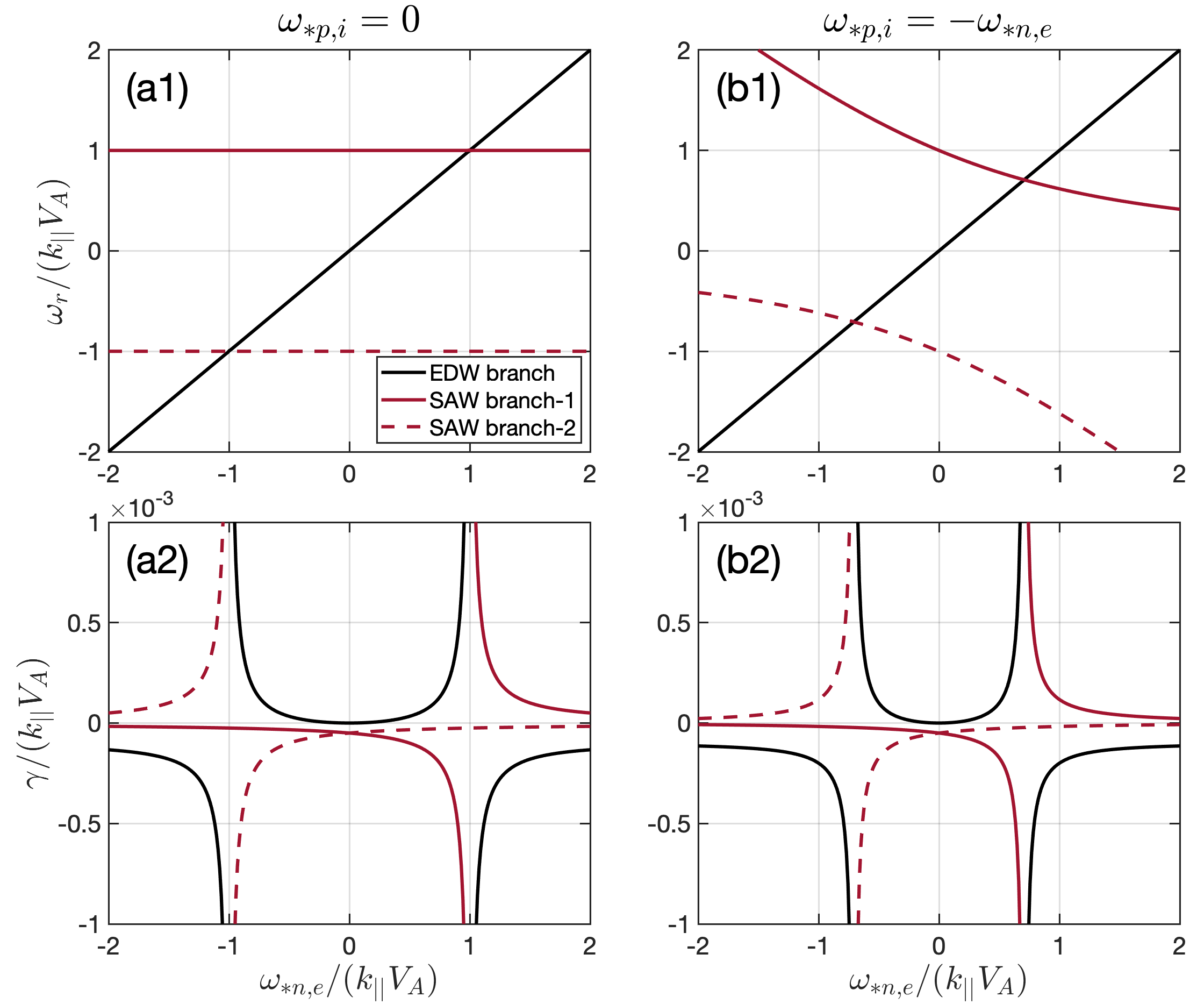}
	\caption{The theoretical dispersion relations of EDW and SAW normal modes by solving Eq. \eqref{DR_DAW2} with $\omega_{*p,i}=0$ and $\omega_{*p,i}=-\omega_{*n,e}$ respectively. Other fixed parameters include $S=10^{6}$ and $k_\perp^2/k_{||}^2 = 100$. The dependences of normalized (a1) real frequency $\hat{\omega}_r=\omega_r/(k_{||}V_A)$ and (a2) growth rate $\hat{\gamma}=\gamma/(k_{||}V_A)$ on normalized EDD frequency $\omega_{*n,e}/(k_{||}V_A)$ in $\omega_{*p,i}=0$ case, and (b1) and (b2) show corresponding results of $\omega_{*p,i}=-\omega_{*n,e}$ case.}
	\label{DAW_theory}	
\end{figure}

First, each normal mode real frequency and growth rate dependences on EDD frequency are studied for two different IDD frequency cases of $\omega_{*p,i}=0$ and $\omega_{*p,i}=-\omega_{*n,e}$ respectively, and other key parameters in Eq.\eqref{DR_DAW2} are fixed as $S = 10^6$ and $k_\perp^2/k_{||}^2=100$. The resistivity effect on EDW/SAW dispersion relation can be checked from $\omega_{*p,i}=0$ case as shown in figures \ref{DAW_theory} (a1) and (a2): the real frequencies of EDW and SAW are barely affected by resistivity, while the EDW and SAW branches become excited or damped in certain $\omega_{*n,e}/(k_{||}V_A)$ regimes, i.e., $\gamma_{EDW}>0$ corresponds to $|\omega_{*n,e}/(k_{||}V_A)|<1$ and $\gamma_{SAW}>0$ corresponds to $|\omega_{*n,e}/(k_{||}V_A)|>1$. Considering tokamak configuration with sheared magnetic field, the dominant unstable normal mode can be either EDW branch away from the rational surface or SAW branch in the vicinity of rational surface. The results of $\omega_{*p,i}=-\omega_{*n,e}$ case are shown in figures \ref{DAW_theory} (b1) and (b2), and it is found that the EDW real frequency is barely affected by $\omega_{*p,i}$, while two SAW real frequency curves are modified significantly being consistent to the expression of Eq. \eqref{DR_DAW} that $\omega_{*p,i}$ leads to a shift on SAW real frequency. Consequently, the intersection point position between EDW and SAW branches is changed due to finite $\omega_{*p,i}$, and the unstable domain in $\omega_{*n,e}/(k_{||}V_A)$ coordinate becomes smaller for EDW while becomes larger for SAW as shown in figure \ref{DAW_theory} (b2). It should be pointed out that finite $\omega_{*n,e}$ is responsible for EDW/SAW destabilization rather than finite $\omega_{*p,i}$, since $\gamma_{EDW} = 0$ and $\gamma_{SAW}<0$ at $\omega_{*n,e}/(k_{||}V_A)=0$ point as shown in figures \ref{DAW_theory} (a2) and (b2).

\begin{figure}[H]
	\center
	\includegraphics[width=0.8\textwidth]{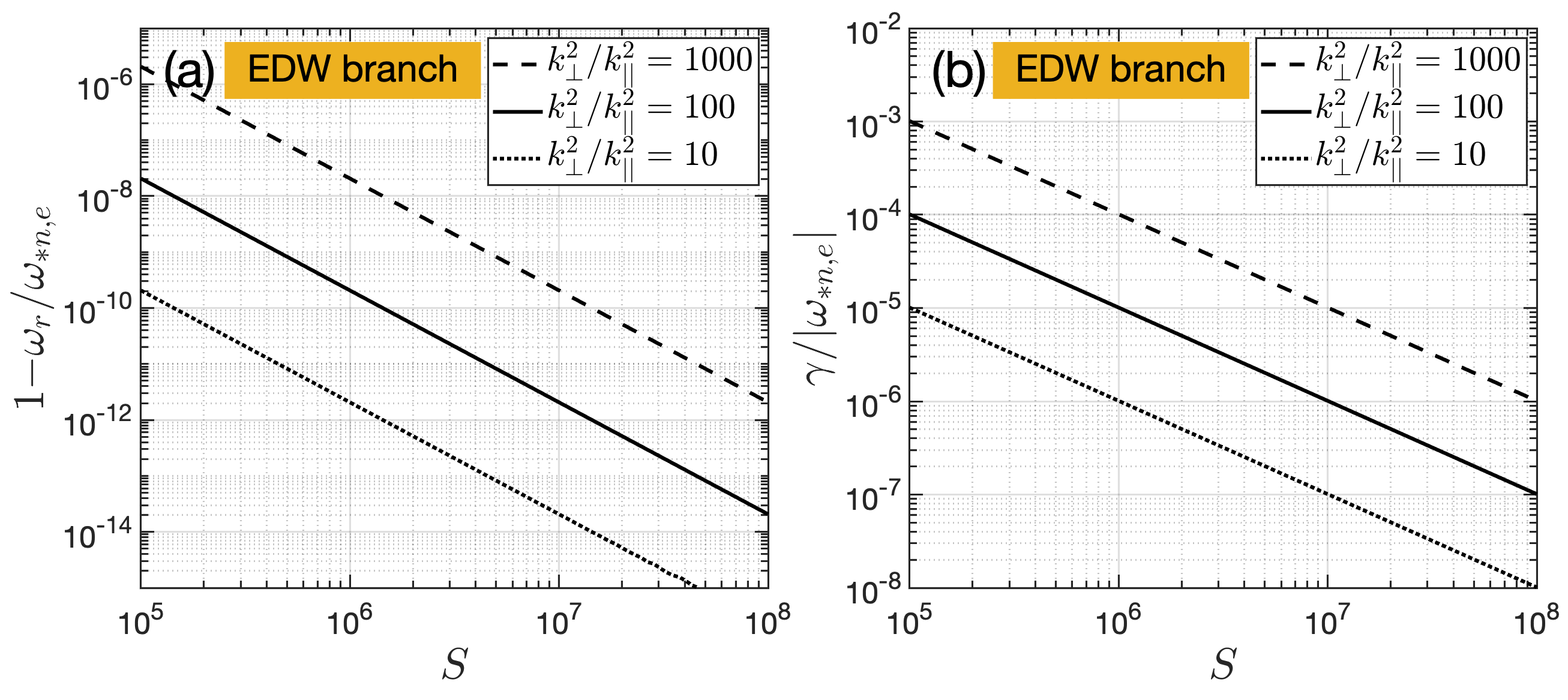}
	\caption{The Lundquist number $S$ scan for unstable EDW branch in Eq. \eqref{DR_DAW2}. (a) Relative frequency difference $1-\omega_r/\omega_{*n,e}$ and (b) normalized growth rate $\hat{\gamma}=\gamma/|\omega_{*n,e}|$. The EDD and IDD frequencies are fixed as $\omega_{*n,e}/(k_{||}V_A)=0.1$ and $\omega_{*p,i} = 0$.}
	\label{DAW_theory3}	
\end{figure}

\begin{figure}[H]
	\center
	\includegraphics[width=0.8\textwidth]{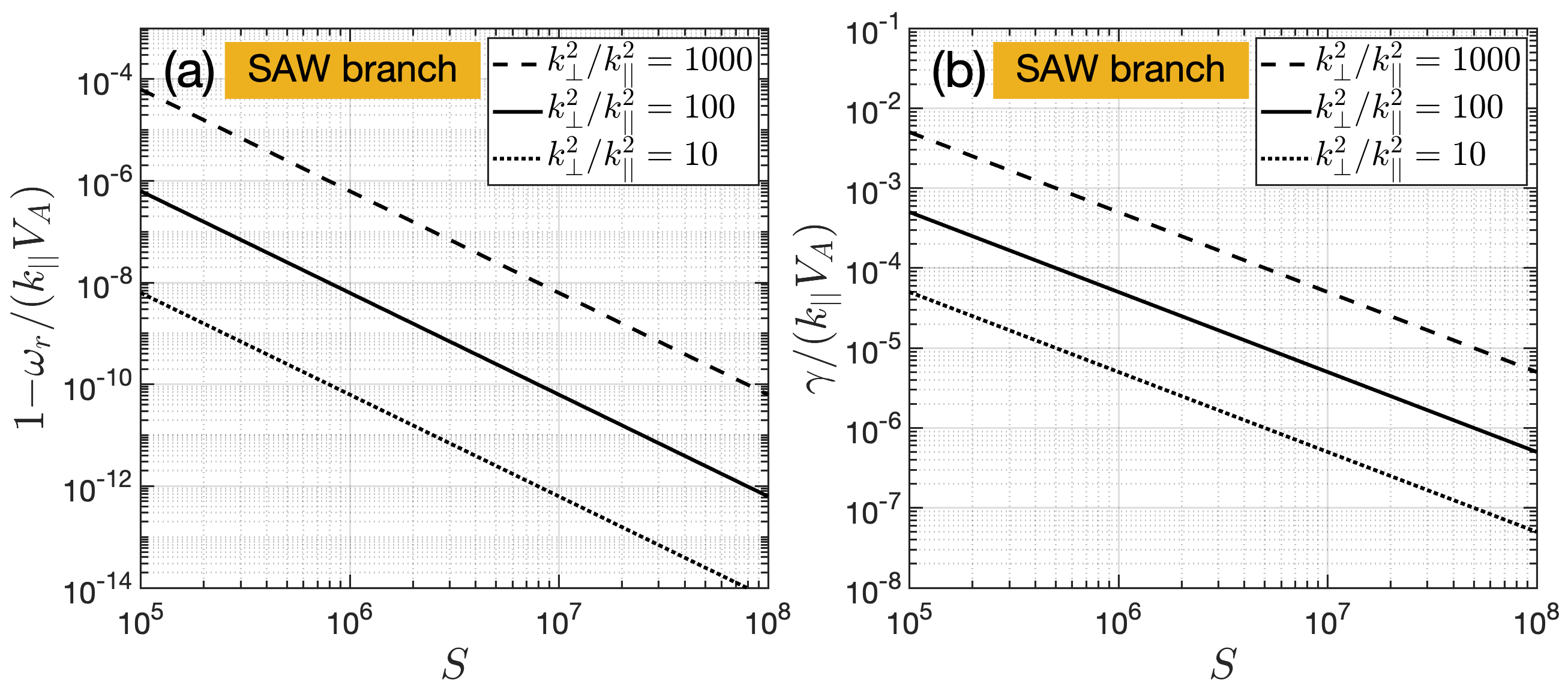}
	\caption{The Lundquist number $S$ scan for unstable SAW branch in Eq. \eqref{DR_DAW2}. (a) Relative frequency difference $1-\omega_r/(k_{||}V_A)$ and (b) normalized growth rate $\hat{\gamma}=\gamma/(k_{||}V_A)$. The EDD and IDD frequencies are fixed as $\omega_{*n,e}/(k_{||}V_A)=2$ and $\omega_{*p,i} = 0$. }
	\label{DAW_theory4}	
\end{figure}

Second, we analyze the dispersion relations of EDW and SAW branches by scanning Lundquist number $S$ as shown in figures \ref{DAW_theory3} and \ref{DAW_theory4}, respectively, where the fixed parameters include $\omega_{*n,e}/(k_{||}V_A)=0.1$ for EDW and $\omega_{*n,e}/(k_{||}V_A)=2$ for SAW that locate in each unstable domain in figure \ref{DAW_theory} (a2), and $\omega_{*p,i}=0$. Three cases with different ratios of $k_\perp^2/k_{||}^2=10,100,1000$ are compared with each other. It is found that (i) the real frequencies of EDW and SAW are not sensitive to $k_\perp^2/ k_{||}^2$ and $S$, which are very close to but slightly smaller than the results in the zero-$\eta_{||}$ limiting of $\omega_r=\omega_{*n,e}$ and $\omega_r=k_{||}V_A$ as shown in figures \ref{DAW_theory3} (a) and \ref{DAW_theory4} (a) respectively; (ii) the grow rates of EDW and SAW both exhibit exponential dependences with $k_\perp^2/ k_{||}^2$ and $S$, which decrease with increasing $S$ or decreasing $k_\perp^2/ k_{||}^2$, since the imaginary term on the RHS of Eq. \eqref{DR_DAW2} is proportional to $(1/S)(k_\perp^2/k_{||}^2)$.

In a short summary, the resistivity enables the release of electron free energy through finite $\omega_{*n,e}$, which leads to the destabilization of EDW/SAW normal mode in certain regime of $\omega_{*n,e}/(k_{||}V_A)$. Since the local dispersion relations of DTM and EDW/SAW can be straightforwardly derived from drift-MHD model in section \ref{drift_MHD}, the spatial cross-scale coupling between DTM and EDW/SAW can be inevitable and important with a global treatment of stability analysis.

   \section{Simulation results}\label{simulation}
In this section, we perform MAS simulations to study DTM linear properties by solving Eqs. \eqref{vor3}-\eqref{dne3} in a global manner, which incorporates DTM and EDW/SAW physics as well as their mutual coupling self-consistently. In all simulations, the two-species plasmas of electron and proton ion with quasi-neutrality condition $Z_in_{i0} =en_{e0}$ are considered, and the cylinder geometry is adopted in order to isolate the effects from geometry complexity. The cylinder length is $l=2\pi R_0$ with $R_0 = 1$m, the axial magnetic field is uniform with $B_0 = 1$T, and the electron and ion temperature profiles (i.e., $T_{e0}$ and $T_{i0}$) are also uniform and vary in amplitude for different diamagnetic drift frequencies. The safety factor profile is described by $q = 1.8 + 0.55\hat\psi + 0.19\hat{\psi}^2$ as shown in figure \ref{scaling_RTM_fig} (a), where $\hat{\psi} = \psi/\psi_W$ is poloidal magnetic flux normalized by the wall value.

	\subsection{Benchmarks for $m/n=2/1$ resistive-tearing mode}\label{benchmark}

\begin{figure}[H]
	\center
	\includegraphics[width=0.8\textwidth]{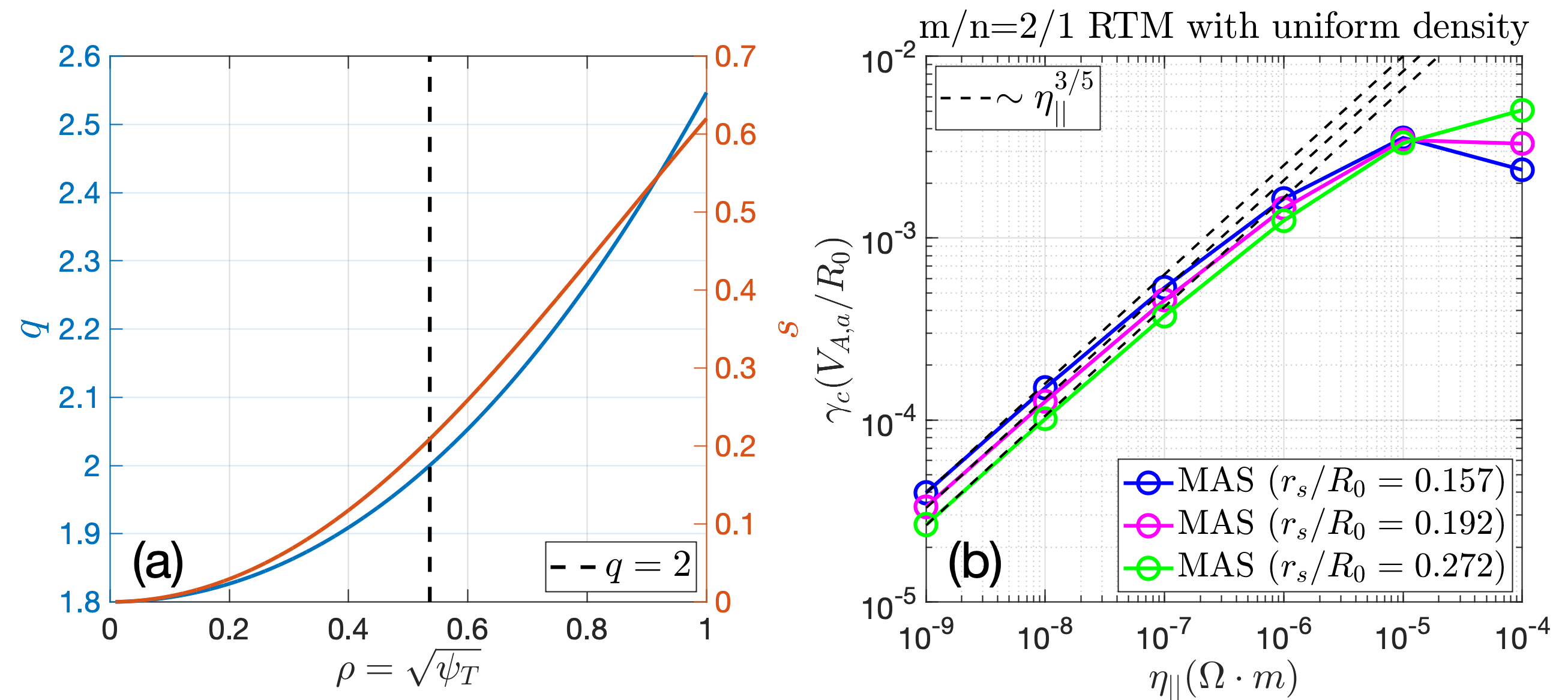}
	\caption{(a) The radial profiles of safety factor $q$ and magnetic shear $s = (r/q)(dq/dr)$. (b) $m/n=2/1$ RTM growth rate dependence on $\eta_{||}$ for different $r_s/R_0$ cases in MAS simulations. The black dashed lines represent theoretical scaling of $\gamma\propto\eta_{||}^{3/5}$ using constant-$\psi$ approximation.}
	\label{scaling_RTM_fig}	
\end{figure}

\begin{figure}[H]
	\center
	\includegraphics[width=1\textwidth]{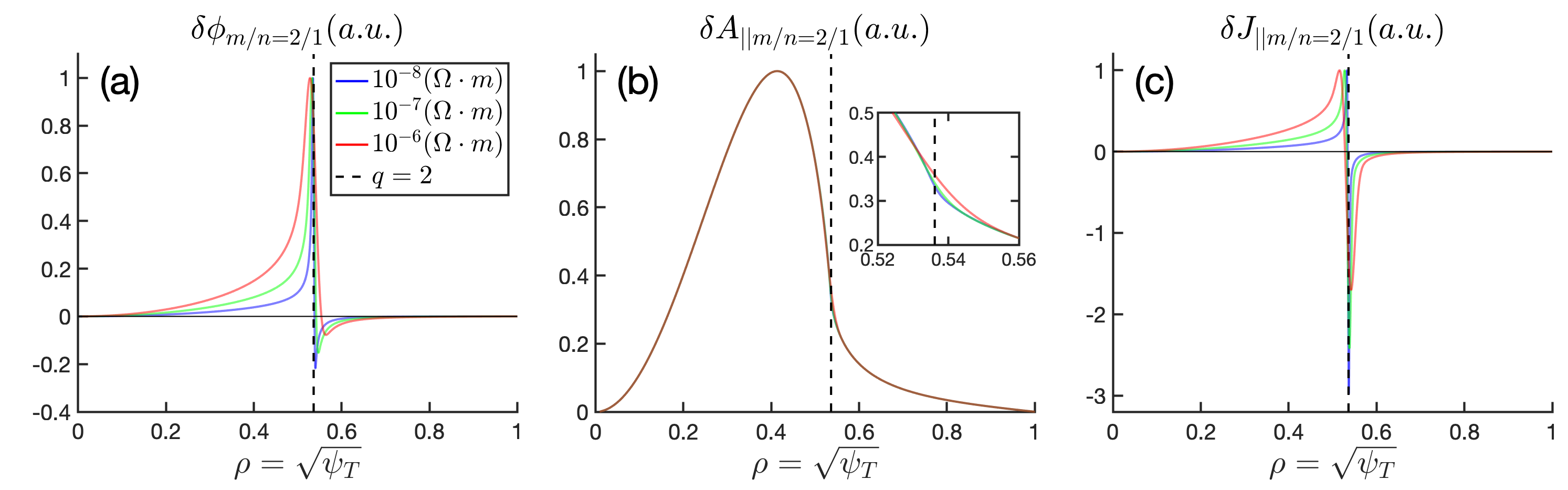}
	\caption{Comparisons of $m=2$ harmonic radial profiles for (a) electrostatic potential $\delta\phi$, (b) parallel vector potential $\delta A_{||}$ and (c) perturbed parallel current $\delta J_{||}$ between different $\eta_{||}$ cases. The normalized distance of $q=2$ rational surface is $r_s/R_0 = 0.272$.}
	\label{RTM_mode_structure_fig}	
\end{figure}

To demonstrate MAS code capability on tearing mode physics, we first verify the $m/n=2/1$ RTM growth rate against the analytic result of Eq. \eqref{DR_RTM}. The ion and electron densities (i.e., $n_{i0}$ and $n_{e0}$) are set to be uniform for benchmark purpose, where the on-axis value is $n_{e0,a} = 10^{14}cm^{-3}$, thus the EDD and IDD terms in Eqs. \eqref{vor3}-\eqref{dne3} are zero and the drift-MHD model reduces to resistive-MHD model being consistent to Eq. \eqref{DR_RTM}. Three series simulations with different inverse aspect ratios, i.e., $r_s/R_0 = 0.157$, $r_s/R_0=0.192$ and $r_s/R_0=0.272$, are carried out for comparison with each other as well as the theoretical scaling indicated by Eq. \eqref{DR_RTM}, where $r_s$ is the distance between $q=2$ rational surface and the magnetic axis. Specifically, the $\eta_{||}$ scan is performed from $10^{-9}\left(\Omega\cdot m\right)$ to $10^{-4}\left(\Omega\cdot m\right)$, and the RTM growth rate $\gamma$ is shown in figure \ref{scaling_RTM_fig} (b). It is seen that MAS simulation results satisfy the theoretical scaling law of $\gamma\propto\eta_{||}^{3/5}$ in the small $\eta_{||}$ regime, while the deviation becomes larger as $\eta_{||}$ increases since constant-$\psi$ approximation is no longer valid due to the broadening of tearing current sheet (i.e, inner region). In figure \ref{RTM_mode_structure_fig}, we compare the RTM mode structures among different $\eta_{||}$ cases with $r_s/R_0=0.272$. The radial structures of electrostatic potential $\delta\phi$ and perturbed parallel current $\delta J_{||}$ are narrowly distributed around the $q=2$ rational surface, and the radial widths can become wider as $\eta_{||}$ increases. In contrast, the parallel vector potential $\delta A_{||}$ has a radially broadening profile with finite amplitude at $q=2$ rational surface, i.e., $|\delta A_{||}(k_{||}=0)|>0$, which is less sensitive to $\eta_{||}$ than $\delta\phi$ and $\delta J_{||}$. So far, the RTM linear properties have been verified by MAS simulations, of which growth rates show good agreements with theory when constant-$\psi$ approximation is valid in small $\eta_{||}$ regime \cite{Furth1963, Furth1973}, and the mode structures well exhibit typical RTM characters.

	\subsection{Global dispersion relation of drift-tearing mode with coupling to electron drift-wave instability}\label{section4_2}

\begin{figure}[H]
	\center
	\includegraphics[width=0.55\textwidth]{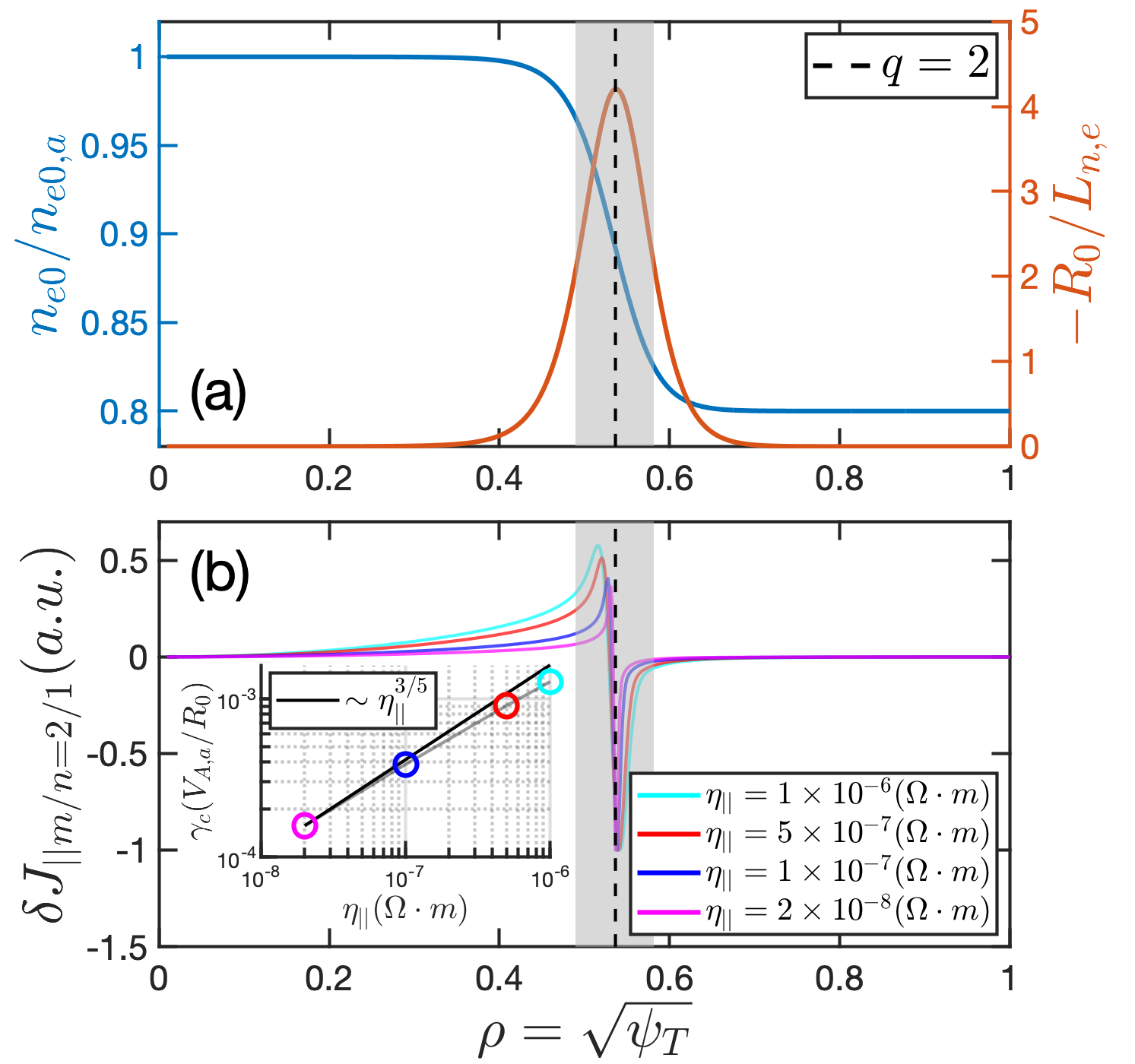}
	\caption{(a) The radial profiles of electron density $n_{e0}$ normalized by on-axis value, and $L_{n,e} = |\nabla ln\left(n_{e0}\right)|^{-1}$ is corresponding scale length. (b) The $\delta J_{||}$ mode structures of $m/n=2/1$ RTM for different $\eta_{||}$ cases. The small panel shows the comparison of RTM growth rate scaling between MAS simulation and theory. The grey shaded region indicates the full width at half maximum amplitude of $-R_0/L_{n,e}$.}
	\label{dtm_ne2_fig}	
\end{figure}

The global nature of DTM is studied in details based on MAS simulations with nonuniform electron density profile, which is described by $n_{e0} = n_{e0,a}\left[1 + 0.1\left(tanh\left(0.32-\hat{\psi}\right)/0.06\right)-1.0\right]$ with on-axis value $n_{e0,a} = 10^{14}cm^{-3}$ as shown in figure \ref{dtm_ne2_fig} (a), and the peak density gradient $-R_0/L_{n,e}\approx 4$ locates at the $q=2$ rational surface, where $L_{n,e} = |\nabla ln\left(n_{e0}\right)|^{-1}$ is the electron density scale length. The ion and electron temperature profiles are uniform but vary in amplitude for different EDD and IDD frequencies. The normalized distance of $q=2$ rational surface is $r_s/R_0=0.272$. Other equilibrium parameters have been introduced in the beginning of this section. 

We first carry out simulations in the zero-temperature limit to examine RTM linear properties within the non-uniform density profile. As shown in figure \ref{dtm_ne2_fig} (b), the radial profiles of $\delta J_{||}$ from different $\eta_{||}$ cases are compared, of which radial width increases with increasing $\eta_{||}$ and most parts of fluctuation are distributed inside the grey shaded region (i.e., full width at half maximum amplitude of $-R_0/L_{n,e}$), and the growth rate scaling with $\eta_{||}$ is close to $\gamma\sim\eta_{||}^{3/5}$. Therefore, the RTM dispersion relation and mode structure from MAS simulations with the proper parameters of nonuniform density and small $\eta_{||}$, almost satisfy the theoretical assumption of constant-$\psi$ approximation, where the radial variations of both density and its gradient (which is proportional to $\omega_{*n,e}$ and $\omega_{*p,i}$) are much slower than tearing current sheet variation. With using finite electron and ion temperatures, one can compare DTM dispersion relation between global MAS simulation and theoretical predication of Eq. \eqref{DR_DTM} (which is based on asymptotic matching method and termed as local theory in the following text) on an equal footing of scale separation.

\begin{figure}[H]
	\center
	\includegraphics[width=0.6\textwidth]{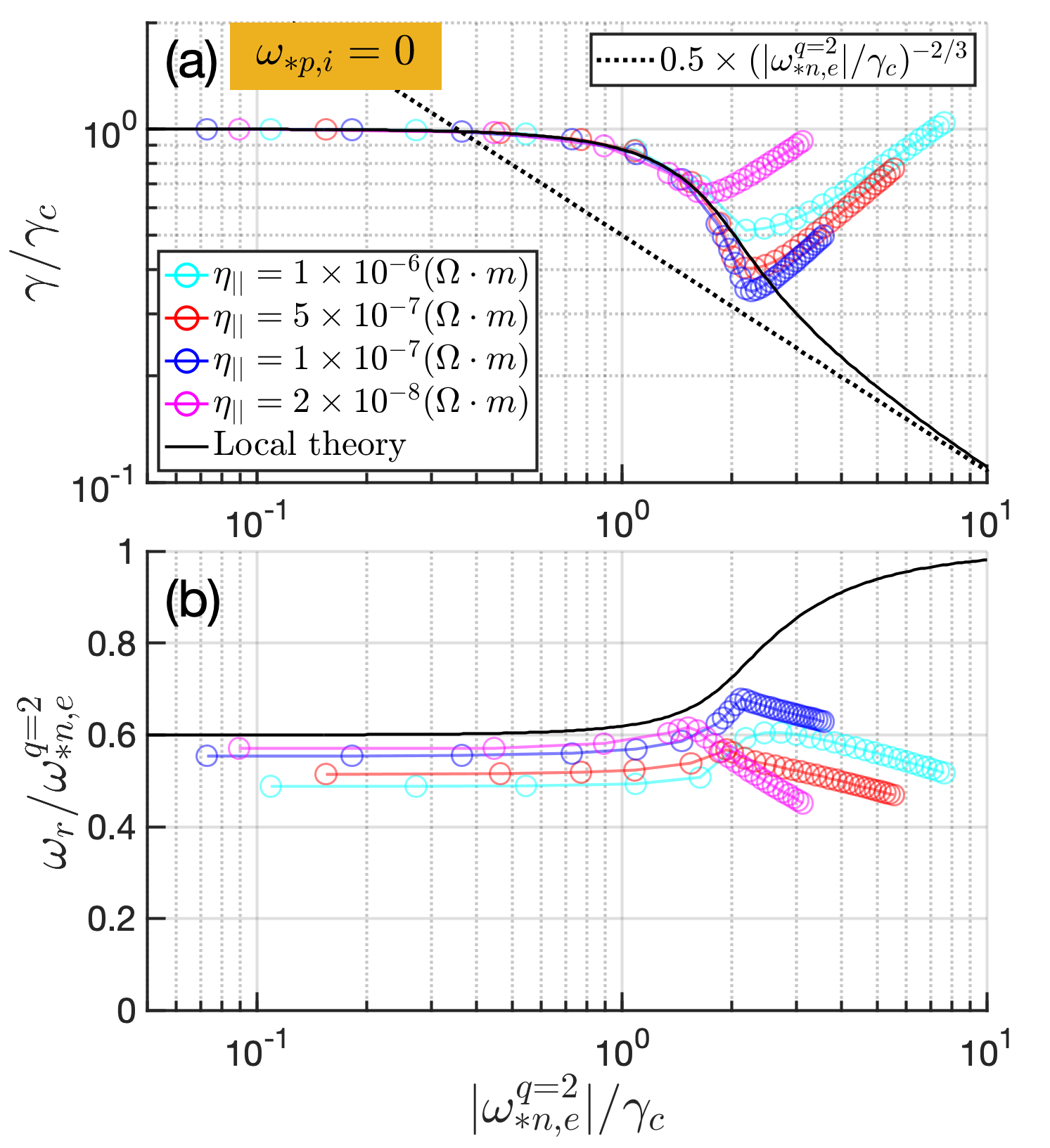}
	\caption{The $m/n=2/1$ DTM (a) growth rate and (b) real frequency dependences on EDD frequency for different $\eta_{||}$ cases. The ion temperature is set to be zero in simulation, i.e., $\omega_{*p,i}=0$. The black solid line represents the local theory result by solving Eq. \eqref{DR_DTM}, which is identical for different $\eta_{||}$ cases due to the label normalization using $\gamma_c$.}
	\label{diff_eta_scan_fig}	
\end{figure}

Building on the identification of RTM dispersion relation with non-uniform electron density, the dependences of DTM growth rate and real frequency on EDD frequency $\omega_{*n,e}$ are studied by scanning electron temperature in MAS, while the ion temperature is set to be zero in this part, since IDD frequency $\omega_{*p,i}$ only shifts DTM growth rate curve in local dispersion relation as shown in figure \ref{DTM_theory_fig}, of which effects on DTM global stability will be analyzed in the end of this work. As shown in figure \ref{diff_eta_scan_fig}, global MAS simulations show the non-monotonic variations of DTM growth rate and real frequency with $|\omega_{*n,e}^{q=2}|/\gamma_c$ (note that $\gamma_c$ is corresponding classical RTM growth rate in the zero-temperature limit, and $\omega_{*n,e}^{q=2}$ represents $\omega_{*n,e}$ value at $q=2$ rational surface), where a turning point can be found in each $\eta_{||}$ case at a critical value of $|\omega_{*n,e}^{crit}|/\gamma_c\sim 2$. Specifically, the DTM growth rate decreases with increasing $|\omega_{*n,e}|$ in the regime of $|\omega_{*n,e}^{q=2}|<|\omega_{*n,e}^{crit}|$ and shows quantitative agreement with local theory, while EDD effect on DTM stability changes to destabilizing rather than stabilizing when $|\omega_{*n,e}^{q=2}|>|\omega_{*n,e}^{crit}|$, where the growth rate increases with increasing $|\omega_{*n,e}|$ that qualitatively deviates from the theoretical scaling of $\gamma\sim |\omega_{*n,e}^{q=2}|^{-2/3}$ in the large limit of $|\omega_{*n,e}^{q=2}|/\gamma_c\gg 1$. The DTM real frequency variation with $|\omega_{*n,e}^{q=2}|/\gamma_c$ is also separated by $\omega_{*n,e}^{crit}$ into two regimes in MAS global simulations. In the regime of $|\omega_{*n,e}^{q=2}|<|\omega_{*n,e}^{crit}|$, the real frequency increases with increasing $|\omega_{*n,e}^{q=2}|/\gamma_c$, which is mostly identical to but slightly smaller than local theory calculation, and the differences can be reduced for small $\eta_{||}$. The reason is that the radial variation of $\omega_{*n,e}$ in global simulation (which corresponds to $-R_0/L_{n,e}$ in figure \ref{dtm_ne2_fig} (a)) can lead to the deviation from theoretical calculation with local $\omega_{*n,e}^{q=2}$, and the DTM current sheet becomes more localized in small $\eta_{||}$ regime so that the local theory becomes more valid. Therefore, both DTM real frequency and growth rate from global MAS simulations differ from local theory qualitatively when EDD frequency amplitude exceeds a certain threshold of $|\omega_{*n,e}^{crit}|\sim 2\gamma_c$, which indicate that global effects in drift-MHD model can not be ignored for DTM in the regime of $|\omega_{*n,e}^{q=2}|>|\omega_{*n,e}^{crit}|$.

\begin{figure}[H]
	\center
	\includegraphics[width=1\textwidth]{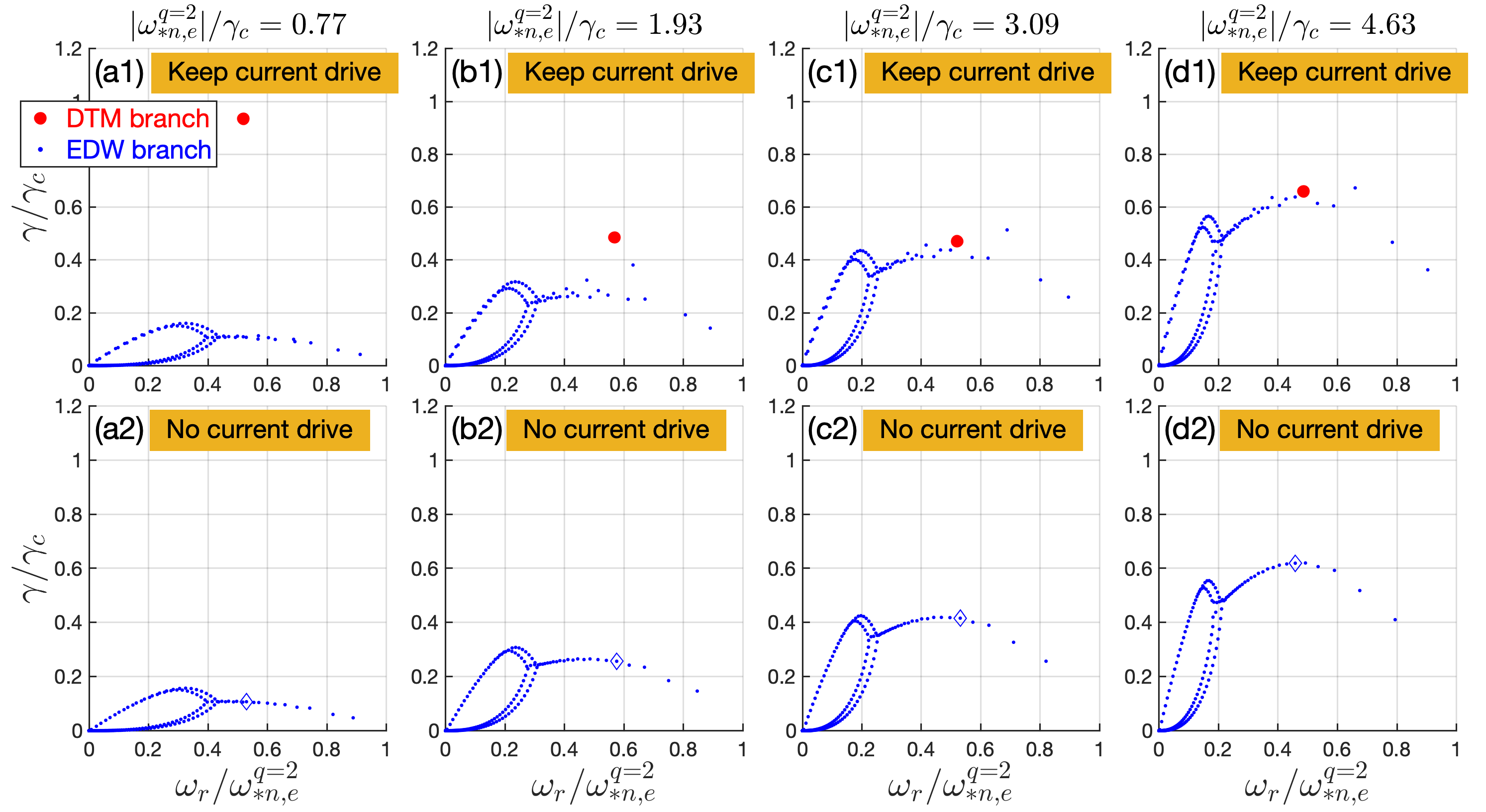}
	\caption{The spectra of unstable mode solutions on $\omega_r-\gamma$ complex plane for different $|\omega_{*n,e}^{q=2}|/\gamma_c$ cases with $\eta_{||} = 5\times 10^{-7}\left(\Omega\cdot m\right)$, the red circle represents DTM solution, and the blue dots represent EDW solutions. The upper and lower rows represent cases with and without current drive term (i.e., $J_{||0}$-related term in Eq. \eqref{vor3}) respectively. The EDW solutions with real frequency being close to corresponding DTMs are marked by blue diamonds in (a2)-(d2).}
	\label{spectrum}	
\end{figure}

To understand the underlying destabilizing mechanism in the regime of $|\omega_{*n,e}^{q=2}|>|\omega_{*n,e}^{crit}|$ from global simulations that is opposite to local theory, we further investigate the distribution of all unstable solutions on $\omega_r-\gamma$ complex plane by taking advantage of the EVP approach in MAS code, which is helpful for identifying each physics branch and revealing their mutual coupling. We choose four different $|\omega_{*n,e}^{q=2}|/\gamma_c$ cases on DTM dispersion relation curve of $\eta_{||}=5\times10^{-7}\left(\Omega\cdot m\right)$ in figure \ref{diff_eta_scan_fig}, and analyze corresponding unstable spectra as shown in figures \ref{spectrum} (a1)-(d1). It is seen that there are only two branches of unstable solutions from global MAS simulations in sheared magnetic field that is less than theoretical predication of three unstable branches in section \ref{theory_EDW_SAW}, the one marked by red circle is identified as DTM, and other solutions marked by blue dots are identified as discretized EDW instabilities, since their growth rates increase with increasing $|\omega_{*n,e}^{q=2}|/\gamma_c$ that are consistent to theoretical analysis of figure \ref{DAW_theory} (a2), i.e., only the unstable EDW normal mode is characterized by the exponential increase on growth rate with increasing $|\omega_{*n,e}|$, while the unstable SAW normal mode exhibits opposite growth rate dependence on $|\omega_{*n,e}|$. Comparing the eigenvalue distributions in figures \ref{spectrum} (a1) and (b1), the DTM growth rate decreases with increasing $|\omega_{*n,e}^{q=2}|/\gamma_c$, which is still larger than all EDW instability solutions in the regime of $|\omega_{*n,e}^{q=2}|<|\omega_{*n,e}^{crit}|\sim 2\gamma_c$. Here, the term `unstable EDW normal mode' refers to the zero-dimensional solution in $(\omega,k)$ space, and the term`discretized EDW instability' refers to radially global solution constrained by boundary condition, which are qualitatively consistent with each other on growth rate and real frequency. However, by comparing figures \ref{spectrum} (c1) and (d1), the DTM growth rate stops decreasing and begins to increase when it encounters the EDW instabilities in the regime of $|\omega_{*n,e}^{q=2}|>|\omega_{*n,e}^{crit}|\sim 2\gamma_c$, which indicates that the coupling between DTM and EDW instabilities leads to DTM destabilization. It should be pointed out that the DTM solutions in figure \ref{diff_eta_scan_fig} and figures \ref{spectrum} (a1)-(d1) are identified by tracing the red circle position from many continuous changed spectra with high resolution in $|\omega_{*n,e}^{q=2}|/\gamma_c$ coordinate, and DTM is not the most unstable mode in figures \ref{spectrum} (c1) and (d1). Moreover, when the current drive effect (i.e., $J_{||0}$-related term in Eq. \eqref{vor3}) is removed, the DTM solution disappears and only the EDW instability solutions remain in figures \ref{spectrum} (a2)-(d2). Comparing figures \ref{spectrum} (d1) and (d2), it is seen that DTM also induces the fluctuations in the solution distribution of EDW instability, which in turn confirms that the strong coupling between DTM and EDW instability is responsible for DTM destabilization in the regime of $|\omega_{*n,e}^{q=2}|>|\omega_{*n,e}^{crit}|\sim 2\gamma_c$.

In addition, the turning point location, namely $|\omega_{*n,e}^{crit}|/\gamma_c$, slightly varies for different $\eta_{||}$ cases in figure \ref{diff_eta_scan_fig}, which is determined by the specific distribution of unstable modes and their mutual coupling. Although SAW can be unstable in the regime of $|\omega_{*n,e}/(k_{||}V_A)|>1$ with considering finite $\eta_{||}$ and $\omega_{*n,e}$ according to theoretical analysis in figure \ref{DAW_theory} (a2), however, destabilization of this SAW branch is very difficult in a sheared magnetic field, due to the narrow unstable domain in the vicinity of rational surface with $k_{||}\sim 0$ and severe continuum damping \cite{Chen2016}, which is not observed in MAS global simulation. Besides the global destabilization of DTM by $\omega_{*n,e}$ due to EDW instability coupling in this work, we note that the diamagnetic drift can also enhance the growth rate of tearing mode in electron-MHD frame work as studied in Ref. \cite{Cai2008}, although the physical mechanisms of these two processes are very different from each other.

\subsection{Mixed mode structure and radial propagation of electron drift-wave instability}\label{section4_3}

\begin{figure}[H]
	\center
	\includegraphics[width=1\textwidth]{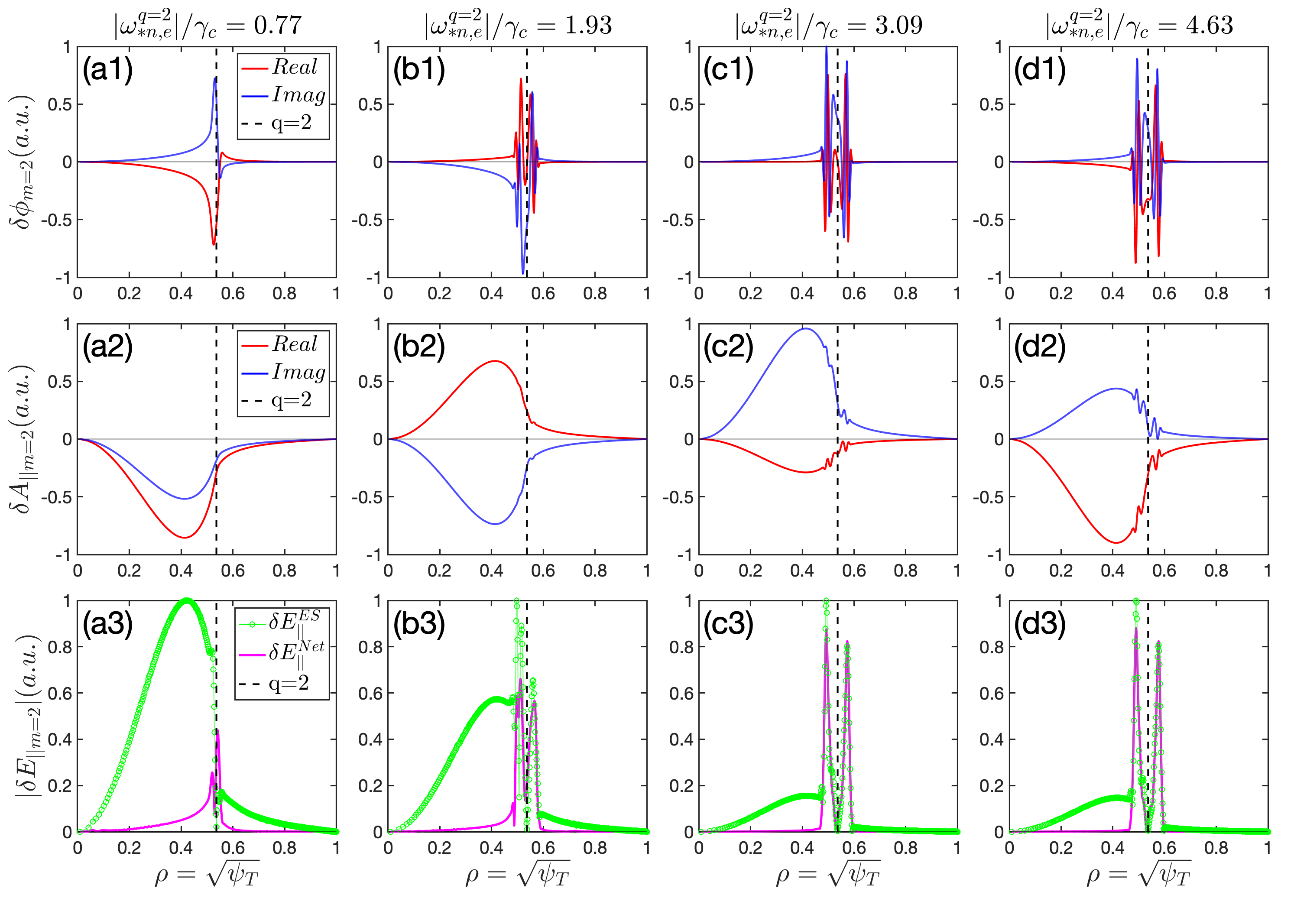}
	\caption{DTM mode structures for different $|\omega_{*n,e}|/\gamma_c$ cases corresponding to red circles in figure \ref{spectrum}. The $m=2$ poloidal harmonic radial plots of (a1)-(d1) electrostatic potential $\delta\phi$, (a2)-(d2) parallel vector potential $\delta A_{||}$, and (a3)-(d3) the parallel electric field $\delta E_{||}$, where the green line with circle represents $\delta E_{||}^{ES} = -\mathbf{b_0}\cdot\nabla\delta\phi$ and the magenta line represents $\delta E_{||}^{Net} = -\mathbf{b_0}\cdot\nabla\delta\phi - (1/c)\partial_t\delta A_{||}$.}
	\label{DTM_structure}	
\end{figure}

Figure \ref{DTM_structure} shows the DTM mode structures of $\delta\phi$ (upper row) and $\delta A_{||}$ (middle row) for different $|\omega_{*n,e}^{q=2}|/\gamma_c$ cases with $\eta_{||}=5\times 10^{-7}\left(\Omega\cdot m\right)$, and the electrostatic component of parallel electric field $\delta E_{||}^{ES} = -\mathbf{b_0}\cdot\nabla\delta\phi$ and net parallel electric field $\delta E_{||}^{Net}= -\mathbf{b_0}\cdot\nabla\delta\phi - (1/c)\partial_t\delta A_{||}$ (bottom row) are also compared to clarify mode polarization \cite{Liu2017,Du2024}, namely, the electrostatic and inductive components mostly cancel each other for Alfv\'enic polarization with $|\delta E_{||}^{Net}|\ll|\delta E_{||}^{ES}|$, while the net parallel electric field is dominated by the electrostatic component for electrostatic polarization with $|\delta E_{||}^{Net}|\approx|\delta E_{||}^{ES}|$. Regarding to resistive drift-MHD model in this work, the Alfv\'enic polarization is intuitive and generally exists in MHD framework, the EDD/IDD and plasma resistivity that break ideal-MHD state can induce the electrostatic polarization. As $|\omega_{*n,e}^{q=2}|/\gamma_c$ increases, the short-wavelength radial oscillations are excited and gradually become dominant on $\delta\phi$ eigenfunctions in figures \ref{DTM_structure} (a1)-(d1), which locate in the domain with finite density gradient. While the $\delta A_{||}$ eigenfunctions in figures \ref{DTM_structure} (a2)-(d2) are much smoother and exhibit the tearing mode structure, which are only slightly modified by short-wavelength radial oscillations. Moreover, by comparing the $\delta E_{||}^{ES}$ and $\delta E_{||}^{Net}$ radial structures in figures \ref{DTM_structure} (a3)-(d3), it is seen that $|\delta E_{||}^{ES}|\gg|\delta E_{||}^{Net}|$ is dominant for the marco-radial-scale structure indicating Alfv\'enic polarization, which corresponds to the tearing mode component. In contrast $|\delta E_{||}^{ES}|\approx|\delta E_{||}^{Net}|$ is dominant for the fine-radial-scale structure (i.e., two peaks on both sides of $q=2$ rational surface) indicating electrostatic polarization, which is consistent with EDW instability character. 



\begin{figure}[H]
	\center
	\includegraphics[width=0.5\textwidth]{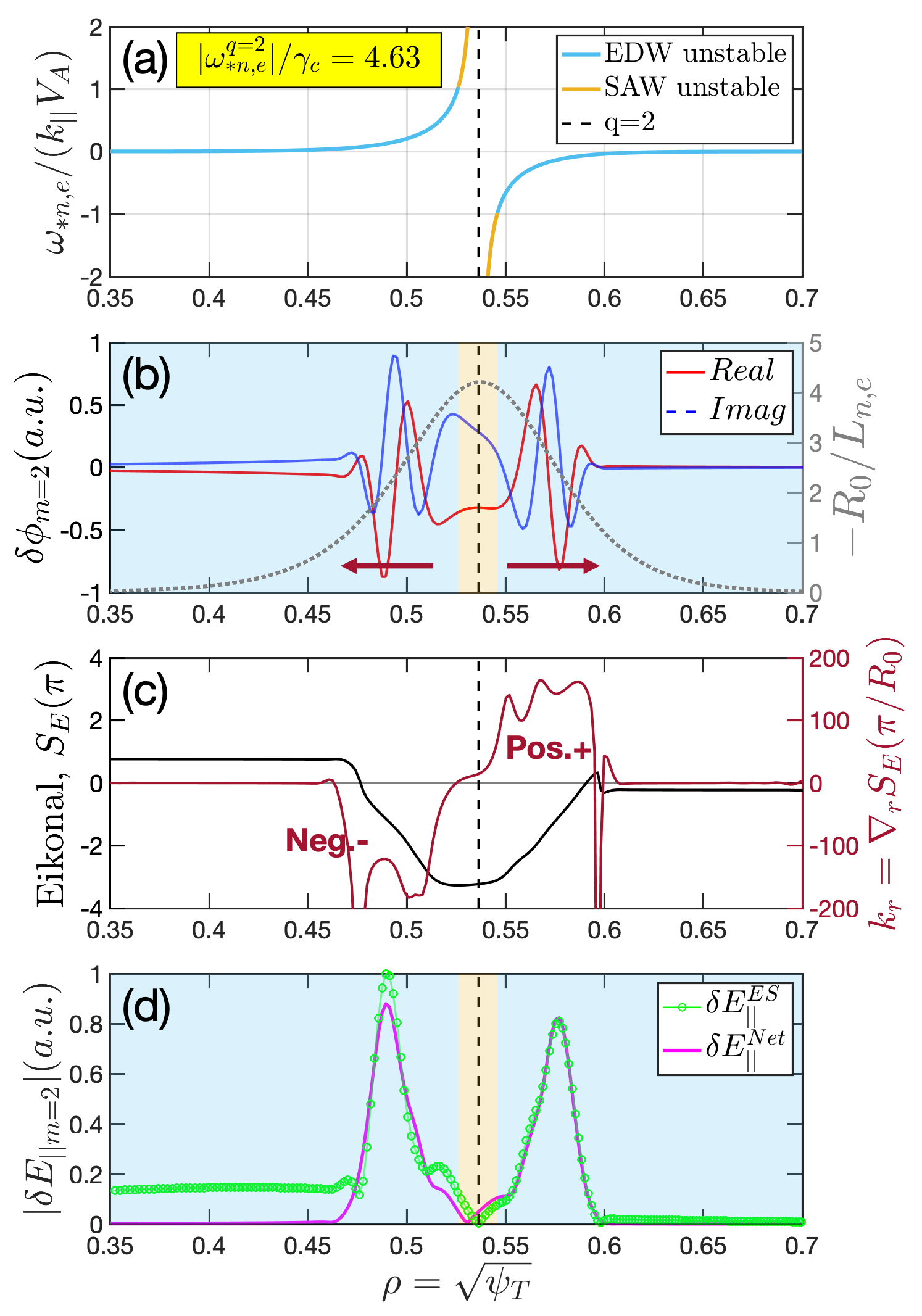}
	\caption{$\delta\phi$ mode structure analysis for $|\omega_{*n,e}^{q=2}|/\gamma_c=4.63$ case in figure \ref{DTM_structure}. The radial profiles of (a) $\omega_{*n,e}/(k_{||}V_A)$ ratio, (b) $m=2$ poloidal harmonic of $\delta\phi$, (c) eikonal $S_E$ and its gradient $\nabla_rS_E$, (d) the amplitudes of $\delta E_{||}^{ES} = -\mathbf{b_0}\cdot\nabla\delta\phi$ and $\delta E_{||}^{Net} = -\mathbf{b_0}\cdot\nabla\delta\phi - (1/c)\partial_t\delta A_{||}$. The arrows represent the radial propagation direction of $\delta\phi$ perturbation. The cyan and yellow shaded regions represent the radial domains of unstable EDW and unstable SAW respectively. }
	\label{DTM_structure_zoom}	
\end{figure}

Next, we focus on $|\omega_{*n,e}^{q=2}|/\gamma_c=4.63$ case and analyze the short-wavelength oscillations on $\delta\phi$ perturbation in detail. In section \ref{theory_EDW_SAW}, we have theoretically shown that the EDW branch is unstable in the regime of $|\omega_{*n,e}/(k_{||}V_A)|<1$ and the SAW branch is unstable in the regime of $|\omega_{*n,e}/(k_{||}V_A)|>1$ for zero-$\omega_{*p,i}$ situation. To further verify with normal mode theory, the radial variation of $\omega_{*n,e}/(k_{||}V_A)$ of this case is shown in figure \ref{DTM_structure_zoom} (a) and compare with the radial distribution of $\delta \phi$ perturbation in figure \ref{DTM_structure_zoom} (b), it is seen that the short-wavelength oscillations are located in the EDW unstable region, which again confirms the EDW instability in addition to the evidences mentioned before, including growth rate increase with $|\omega_{*n,e}|$ in figure \ref{spectrum} and electrostatic polarization in figure \ref{DTM_structure} (d3) and figure \ref{DTM_structure_zoom} (d). Meanwhile, the radial wavelength of EDW instability (i.e., short-wavelength oscillation) is much shorter than the radial scale of $-R_0/L_{n,e}$ profile as shown in figure \ref{DTM_structure_zoom} (b), thus the eikonal approximation can be applied for describing the wave-packet propagation in non-uniform plasmas \cite{Weinberg1962,Lu2012}, and the $\delta\phi$ perturbation can be expressed using eikonal form as
\begin{flalign}\label{}
	\begin{split}
		\delta\phi_{m=2}(\mathbf{r}) = A(\mathbf{r})e^{iS_E(\mathbf{r})}
	\end{split},
\end{flalign}
where $A(\mathbf{r})$ is the wave amplitude and $S_E(\mathbf{r})$ is the eikonal related to the radial wave vector
\begin{flalign}\label{}
	\begin{split}
		\mathbf{k_r}=\frac{\partial S_E}{\partial \mathbf{r}}
	\end{split}.
\end{flalign}

The radial profiles of eikonal $S_E$ and its gradient $\partial S_E/\partial r$ are shown in figure \ref{DTM_structure_zoom} (c). It is seen $\partial S_E/\partial r<0$ and $\partial S_E/\partial r>0$ on the LHS and RHS of $q=2$ rational surface respectively, which indicate the outward propagations of EDW wave-packets from large density gradient domain, and is consistent to early theory on convective instability in a sheared magnetic field \cite{Coppi1966}. We also note that in former study of DTM linear stability by Grasso et al in Ref. \cite{Grasso2001} based on initial value simulation, it is found that the localized mode doesn't exist at high $|\omega_{*n,e}|/\gamma_c$ that exceeds a critical threshold, instead the delocalization of eigenfunction is observed that exhibits radial oscillation as a propagating wave-packet. The fine-radial-scale $\delta\phi$ mode structure arising from EDW instability coupling in this work is consistent to the drift-tearing wave-packet solution in Ref. \cite{Grasso2001}.

\begin{figure}[H]
	\center
	\includegraphics[width=1\textwidth]{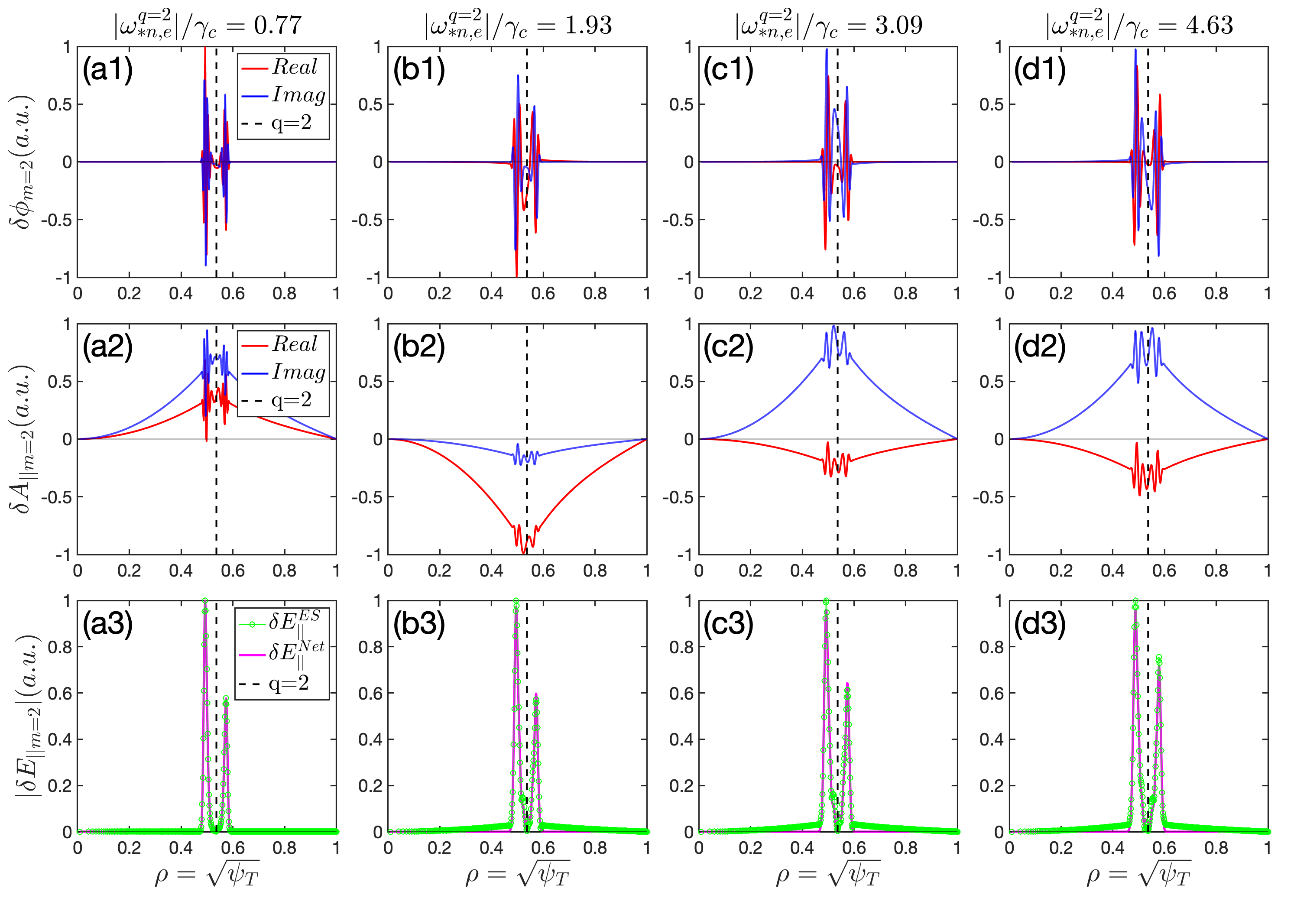}
	\caption{Pure EDW instability mode structures for different $|\omega_{*n,e}|/\gamma_c$ cases corresponding to blue diamonds in figure \ref{spectrum}. Other details are the same with figure \ref{DTM_structure}.}
	\label{EDW_structure}	
\end{figure}

As mentioned before, the EDW instability can also be affected by DTM through the equilibrium current, which can be seen by comparing the blue dots distribution between cases with and without keeping current drive effect (i.e., $J_{||0}$-related term in Eq. \eqref{vor3}) in figure \ref{spectrum}. To demonstrate the role of equilibrium current on short-wavelength oscillation, we show the mode structures of pure EDW instability without current drive effect in figure \ref{EDW_structure}, which correspond to the blue diamonds in figures \ref{spectrum} (a2)-(d2) with real frequencies being close to DTMs. According to figure \ref{EDW_structure}, the characteristics of pure EDW instability mode structure include: (i) $\delta\phi$ perturbation only has high-$k_r$ component, and $\delta A_{||}$ perturbation has both high-$k_r$ and low-$k_r$ components with comparable amplitudes. (ii) The mode polarization is dominantly electrostatic with $|\delta E_{||}^{Net}|\approx|\delta E_{||}^{ES}|$, which indicates $\delta \phi$ perturbation is much more important than $\delta A_{||}$ perturbation.  (iii) $\delta\phi$ perturbation exhibits more singular structure for the small $|\omega_{*n,e}^{q=2}|/\gamma_c$ cases. (iv) $\delta A_{||}$ perturbation no longer exhibits any tearing mode structure. By comparing figures \ref{DTM_structure} and \ref{EDW_structure}, it is seen that the $\delta\phi$ perturbations of DTM and pure EDW instability show similar fine-radial-scale structures for cases with $|\omega_{*n,e}^{q=2}|/\gamma_c>2$ where the influences of equilibrium current are ignorable, while only DTM $\delta A_{||}$ perturbation shows typical tearing mode structure attributing to equilibrium current. 

In a short summary, we identify the roles of equilibrium current drive and density gradient drive on global DTM mode structure in the regime of $|\omega_{*n,e}^{q=2}|>|\omega_{*n,e}^{crit}|$, namely, nearly all $\delta\phi$ perturbation is carried by short-wavelength EDW instability with radially outward propagations through density gradient drive, and most $\delta A_{||}$ perturbation is carried by macroscopic tearing mode through equilibrium current drive, which together consist of a cross-scale mixed mode structure with both electrostatic and Alfv\'enic polarizations.

\subsection{Ion diamagnetic drift effects on global stability of drift-tearing mode}
\begin{figure}[H]
	\center
	\includegraphics[width=0.6\textwidth]{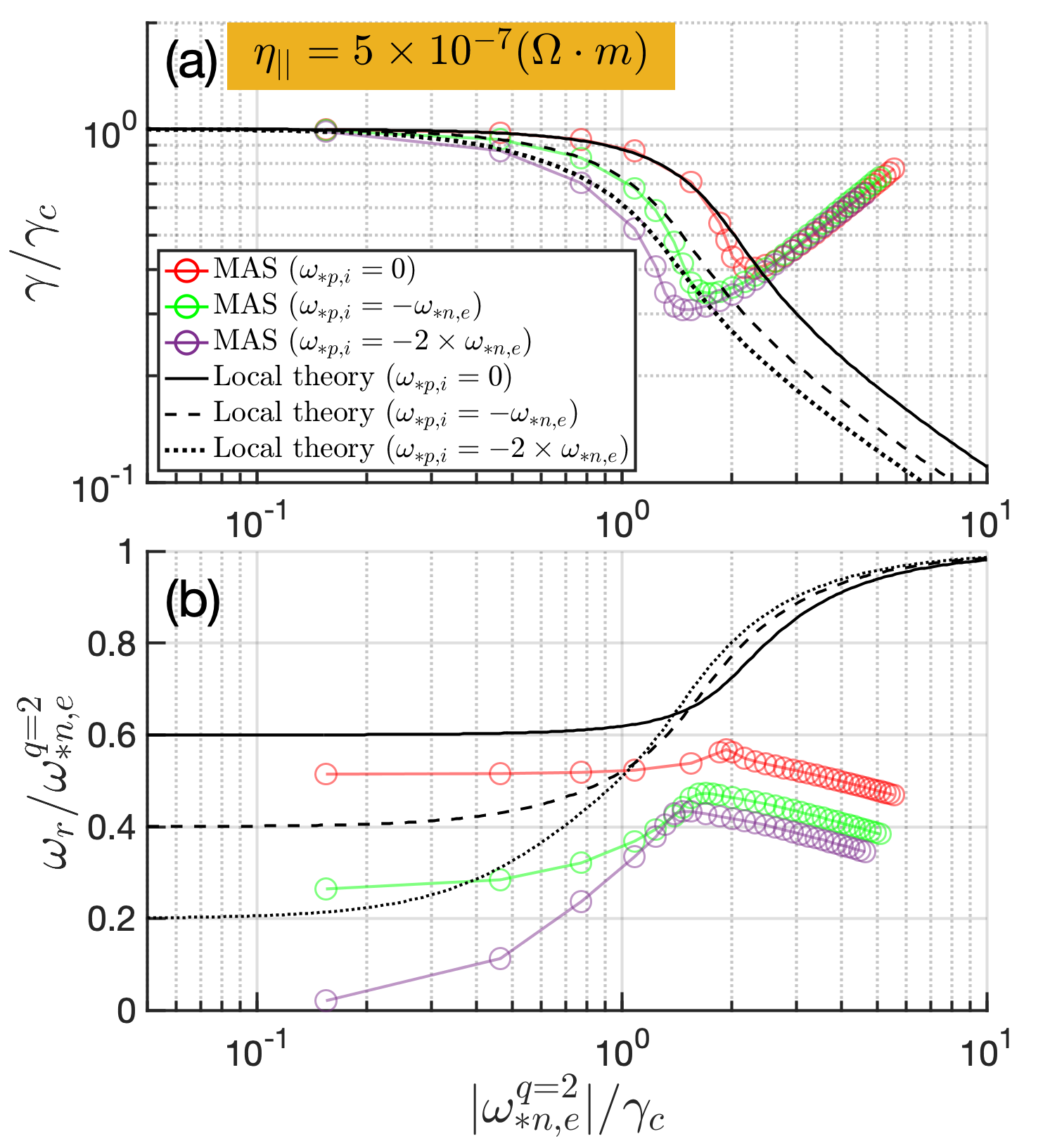}
	\caption{Comparison of $m/n=2/1$ DTM dispersion relation between different $\omega_{*p,i}$ cases. (a) The growth rate and (b) real frequency dependences on EDD frequency. The plasma resistivity is $\eta_{||} = 5\times 10^{-7}(\Omega\cdot m)$. The black solid, dashed and dotted lines represent corresponding local theory results by solving Eq. \eqref{DR_DTM}.}
	\label{diff_eta_scan_fig2}	
\end{figure}

For completeness of this work, we further carry out DTM simulations with finite $\omega_{*p,i}$. The resistivity is $\eta_{||}=5\times10^{-7}\left(\Omega\cdot m\right)$, and both theory and simulation results of DTM growth rate and real frequency for $\omega_{*p,i}=0$, $\omega_{*p,i}=-\omega_{*n,e}$ and $\omega_{*p,i}=-2\omega_{*n,e}$ cases, are compared in figure \ref{diff_eta_scan_fig2}. Compared to the local theory result that finite $\omega_{*p,i}$ induces an overall stabilizing effect on DTM as discussed in section \ref{theory_DTM}, $\omega_{*p,i}$ effects on global DTM stability become more complex in MAS simulation: (i) Finite $\omega_{*p,i}$ not only shifts the DTM growth rate curve towards small $|\omega_{*n,e}|/\gamma_c$ regime, but also leads to the late occurrence of destabilizing turning point. (ii) In the regime of $|\omega_{*n,e}^{q=2}|<|\omega_{*n,e}^{crit}|$, finite $\omega_{*p,i}$ leads to DTM stabilization in agreement with local theory. In the regime of $|\omega_{*n,e}^{q=2}|>|\omega_{*n,e}^{crit}|$, the DTM growth rate curves from different $\omega_{*p,i}$ cases overlap with each other in their common $|\omega_{*n,e}^{q=2}|/\gamma_c$ domains of destabilization, which implies DTM growth rate is insensitive to $\omega_{*p,i}$ in this regime. (iii) Finite $\omega_{*p,i}$ generally induces the DTM real frequency to shift towards IDD direction, and a similar phenomenon of frequency upshift in IDD direction is also found in the study of Alfv\'en continua and Alfv\'en eigenmodes in tokamaks \cite{Bao2020}.

\begin{figure}[H]
	\center
	\includegraphics[width=0.6\textwidth]{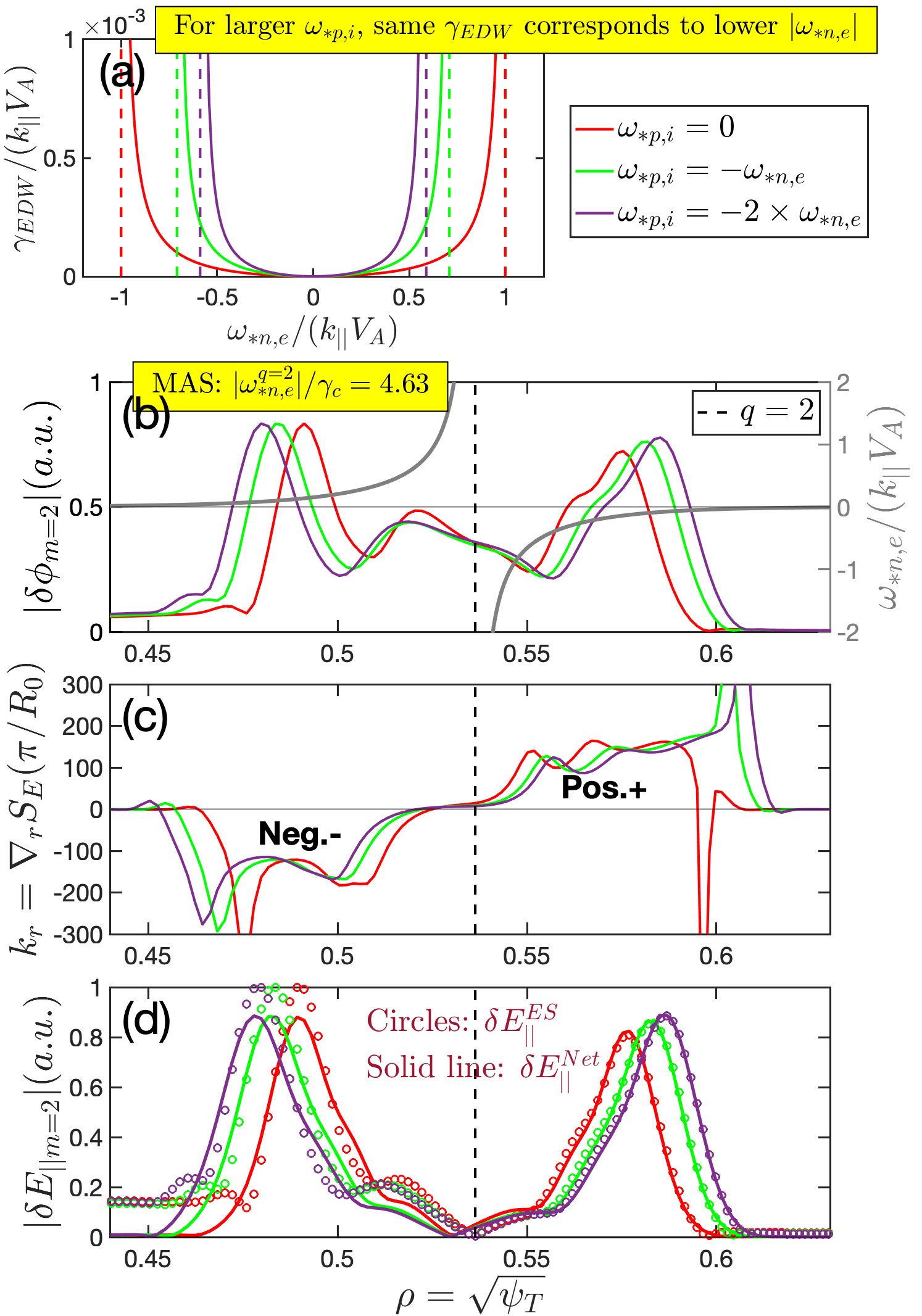}
	\caption{Comparison of DTM mode structures between different $\omega_{*p,i}$ cases with $|\omega_{*n,e}^{q=2}|/\gamma_c=4.63$. (a) Theoretical analysis of EDW instability growth rate using Eq. \eqref{DR_DAW2}. MAS simulations of the radial profiles of (b) $\delta\phi$ amplitude (i.e., $|\delta\phi_{m=2}|$), (c) gradient of eikonal $\nabla_r S_E$, and (d) amplitudes of $\delta E_{||}^{ES} = -\mathbf{b_0}\cdot\nabla\delta\phi$ and $\delta E_{||}^{Net} = -\mathbf{b_0}\cdot\nabla\delta\phi - (1/c)\partial_t\delta A_{||}$.}
	\label{IDD_mode_structure}	
\end{figure}

To understand the reason why DTM growth rates are almost the same for different $\omega_{*p,i}$ cases in the regime of $|\omega_{*n,e}^{q=2}|>|\omega_{*n,e}^{crit}|$ that overlap with each other, we focus on how $\omega_{*p,i}$ affects the dispersion relation of unstable EDW normal mode in theory and corresponding $\delta\phi$ mode structure in simulation, of which coupling effects are crucial to DTM stability in this regime as clarified in sections \ref{section4_2} and \ref{section4_3}. According to Eq. \eqref{DR_DAW2}, we compare the unstable EDW normal mode growth rate between different $\omega_{*p,i}$ cases by scanning $\omega_{*n,e}/(k_{||}V_A)$ as shown in figure \ref{IDD_mode_structure} (a) (the calculation parameters are the same with figure \ref{DAW_theory}), the color-dashed lines indicate the $\omega_{*n,e}/(k_{||}V_A)$ limits of unstable EDW normal mode, which are only determined by $\omega_{*p,i}$ and independent of Lundquist number $S$ and $k_\perp^2/k_{||}^2$ ratio. It is seen that for larger $\omega_{*p,i}$ case, the same growth rate of EDW normal mode corresponds to the lower value of $|\omega_{*n,e}/(k_{||}V_A)|$. Next, we compare the MAS simulation results of $|\delta\phi_{m=2}|$ radial profiles between different $\omega_{*p,i}$ cases with $|\omega_{*n,e}|/\gamma_c=4.63$ in figure \ref{IDD_mode_structure} (b). In all $\omega_{*p,i}$ cases, $\delta\phi$ fluctuations mainly distribute in the radial domain of $|\omega_{*n,e}/(k_{||}V_A)|\ll 1$ that well satisfy the constraints of unstable EDW (i.e., below the limits marked by the color-dashed lines in figure \ref{IDD_mode_structure} (a)). As $\omega_{*p,i}$ increases, it is found that $\delta\phi$ fluctuations move towards smaller $|\omega_{*n,e}/(k_{||}V_A)|$ domain in figure \ref{IDD_mode_structure} (b) being consistent with the conclusion of theoretical analysis in figure \ref{IDD_mode_structure} (a), which implies that the density gradient drive of each species becomes weaker and cancels the new contribution from ion channel, therefore the grow rate of DTM with strong EDW instability coupling effect is insensitive to $\omega_{*p,i}$, due to the fact that $\omega_{*p,i}$ effect induces the shift of $\delta\phi$ mode structure towards small density gradient domain. Figures \ref{IDD_mode_structure} (c) and (d) show the radial profiles of $\nabla_rS_E$ and $\delta E_{||}$ for different $\omega_{*p,i}$ cases, and finite $\omega_{*p,i}$ effect doesn't induce any qualitative differences from zero-$\omega_{*p,i}$ results, i.e., the outward propagation of shot-wavelength oscillation and electrostatic polarization still persist.

\begin{figure}[H]
	\center
	\includegraphics[width=0.6\textwidth]{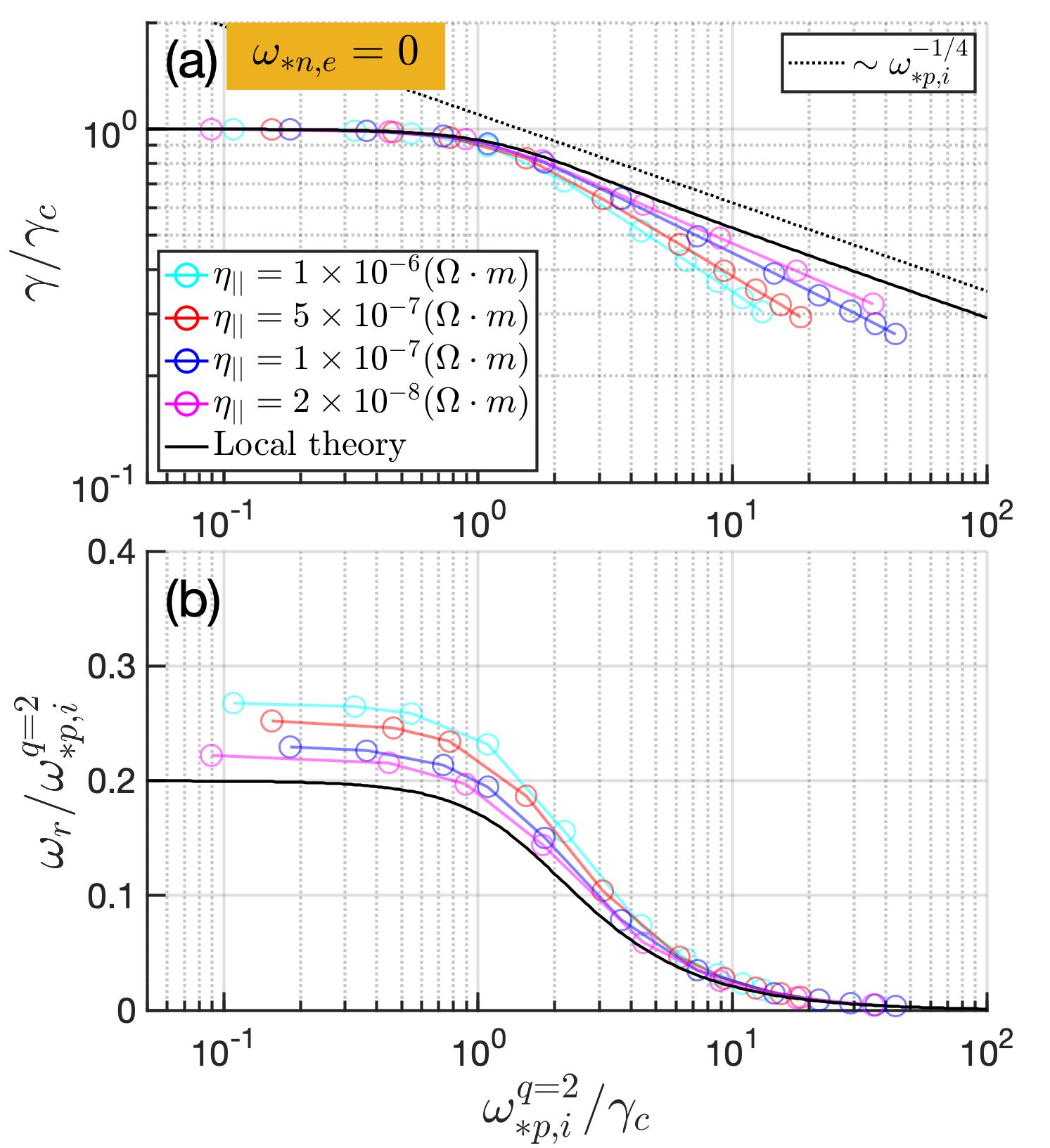}
	\caption{The $m/n=2/1$ RTM (a) growth rate and (b) real frequency dependences on IDD frequency for different $\eta_{||}$ cases. The electron temperature is set to be zero in simulation, i.e., $\omega_{*n,e}=0$. The black solid line represents the local theory result by solving Eq. \eqref{DR_DTM}, which is identical for different $\eta_{||}$ cases due to the label normalization using $\gamma_c$.}
	\label{diff_eta_scan_fig3}	
\end{figure}

Finally, the single $\omega_{*p,i}$ effect is investigated by setting $\omega_{*n,e}=0$, which is no longer the condition for DTM regime and returns to RTM. The dependences of RTM growth rate and real frequency on $\omega_{*p,i}$ for different $\eta_{||}$ cases are shown in figure \ref{diff_eta_scan_fig3}. It is seen that the $\omega_{*p,i}$ only causes RTM stabilization in both theory and simulation, and the local theory Eq. \eqref{DR_DTM} predicts the RTM growth rate scaling with $\omega_{*p,i}$ as $\gamma\sim\omega_{*p,i}^{-1/4}$ in the large limit of $\omega_{*p,i}/\gamma_c\gg 1$. Note that MAS global simulation results slightly deviate from local theory predication as $\eta_{||}$ increases, which attribute to both $\omega_{*p,i}$ nonuniformity and the constant-$\psi$ approximation that break the consistency between simulation and theory, and the thin current sheet in small $\eta_{||}$ simulation case can better satisfy the local theory assumption.

	\section{Conclusion and discussion}
	
In this work, the linear properties of DTM have been investigated by comparing results between asymptotic theory and global MAS simulations, which emphasize the global characters on dispersion relation, mode structure and polarization. A drift-MHD physics model is applied with keeping the necessary EDD and IDD frequencies on top of resistive reduced-MHD equations, while excludes other higher order effects on tearing mode. The main results are summarized as follows.

\begin{enumerate}
	
	\item \textbf{Uncoupled DTM and EDW/SAW normal modes in local theory analysis.} Based on drift-MHD model equations, two independent dispersion relations are derived for discretized DTM eigenmode and EDW/SAW normal modes respectively, which give following theoretical results: (i) DTM growth rate decreases with increasing $|\omega_{*n,e}|$ with a scaling of $\gamma\sim|\omega_{*n,e}|^{-2/3}$ in the limit of $|\omega_{*n,e}|/\gamma_c \gg 1$. (ii) DTM real frequency is determined by both IDD and EDD frequencies as $\omega_r=\left(3\omega_{*n,e}+\omega_{*p,i}\right)/5$ in the limit of $|\omega_{*n,e}|/\gamma_c \ll 1$, while it gradually transit to $\omega_r=\omega_{*n,e}$ in the limit of $|\omega_{*n,e}|/\gamma_c \gg 1$ that only relies on EDD frequency. (iii) Due to the synergy of finite resistivity and EDD, two normal modes, i.e., EDW and SAW, are coupled with each other and can be destabilized in certain $|\omega_{*n,e}/(k_{||}V_A)|$ regime. (iv) In general, with fixed Lundquist number $S$, $k_\perp^2/k_{||}^2$ and $\omega_{*p,i}$, the unstable EDW growth rate exponentially increases with increasing $|\omega_{*n,e}/(k_{||}V_A)|$, while the unstable SAW growth rate exponentially decreases with increasing $|\omega_{*n,e}/(k_{||}V_A)|$. 
	
	\item \textbf{Global characteristics of DTM in MAS simulation with coupling to EDW instability.} MAS self-consistently treats the linear couplings between DTM and co-existing unstable normal modes, and the dispersion relation and mode structure of the most important solution in DTM branch exhibit global characteristics, which are qualitatively different from local theory results in the limit of $|\omega_{*n,e}|/\gamma_c\gg 1$ that is typical for present-day tokamak and future fusion reactor. There are three main findings: (i) A turning point occurs for DTM growth rate and real frequency variations with EDD frequency, which corresponds to a critical threshold of $\omega_{*n,e}^{crit}$. In the regime of $|\omega_{*n,e}|<|\omega_{*n,e}^{crit}|$, DTM growth rate decreases with increasing $|\omega_{*n,e}|$ that quantitatively agrees with local theory as mentioned above. In the regime of $|\omega_{*n,e}|>|\omega_{*n,e}^{crit}|$, the strong coupling between DTM and EDW instability happens and changes $|\omega_{*n,e}|$ effect on DTM growth rate from stabilizing to destabilizing that qualitatively deviates from local theory, which attribute to the enhanced drive from free energies in both equilibrium current and density gradient. (ii) A cross-scale mixed mode structure forms in the regime of $|\omega_{*n,e}|>|\omega_{*n,e}^{crit}|$ with both electrostatic and Alfv\'enic polarizations, $\delta\phi$ perturbation is dominated by short-wavelength oscillation from EDW instability that is nonlocalized and propagates outwards from the large density gradient domain, while $\delta A_{||}$ perturbation is mostly characterized by macroscopic tearing mode structure. (iii) DTM growth rate is insensitive to $\omega_{*p,i}$ in the limit of $|\omega_{*n,e}|/\gamma_c\gg 1$, since finite $\omega_{*p,i}$ leads to the shift of DTM mode structure towards small density gradient domain, which cancels the extra drive from ion channel. In the zero-$\omega_{*n,e}$ limit, single $\omega_{*p,i}$ effect induces the decrease of RTM growth rate with scaling of $\gamma\sim\omega_{*p,i}^{-1/4}$ that shows agreement between theory and simulation, which is no longer the DTM regime.
	
\end{enumerate}

We note that the nonlocalized DTM mode structure is also found by initial value simulations with slab geometry when $|\omega_{*n,e}|/\gamma_c$  is sufficiently large that exceeds a critical threshold \cite{Grasso2001,Porcelli2001}, which is considered as the drift-tearing wave-packet solution that propagates outwards from the density gradient drive domain with spatial amplification. Compared to former study, new progresses have been made in this work by taking advantage of EVP approach in MAS code, i.e., we calculate all unstable solutions and give eigenvalue distribution on $\omega_r-\gamma$ complex plane, which enables the identification of multiple branches of eigenstates and their mutual couplings. Though the EDD effect has been considered in the theoretical dispersion relation of DTM by Eq. \eqref{DR_DTM}, there are also many discretized EDW instabilities that can be simultaneously excited with short-wavelength oscillations. The underlying physics mechanism of nonlocalized mode structure in the regime of $|\omega_{*n,e}|>|\omega_{*n,e}^{q=2}|$, can be explained as the formation of a new hybrid mode through linear coupling between DTM and EDW instability confirmed by MAS simulation treating various physics effects globally on an equal footing, which exhibits features of EDW instability on $\delta\phi$ perturbation and tearing mode on $\delta A_{||}$ perturbation. 

The minimal model using incompressible assumption and cylinder geometry is applied in this work for demonstrating the global nature of DTM. The full-MHD effects including plasma compressibility and pressure-drive associated with magnetic curvature, are shown to have significant influences on current-driven MHD instability in experimental geometry \cite{Brochard2022}, which will be incorporated in our future study.

	\section{Acknowledgments}
	
	This work is supported by National Natural Science Foundation of China under Grant Nos. 12275351, 12025508, 11835016 and 11905290; the Strategic Priority Research Program of Chinese Academy of Sciences under Grant No. XDB0500302; and the start-up funding of Institute of Physics, Chinese Academy of Sciences under Grant No. Y9K5011R21.

	\appendix
	
	\section{Dimensionless form and numerical scheme}\label{app_A1}
	Applying the Fourier transform $\partial_t \to -i\omega$ and normalizations of $\left(\frac{V_{Ap}}{R_0}t, R_0\nabla,\frac{B_0}{B_a},\frac{n}{n_{ea}},\frac{T}{T_{ea}}\right)\to\left(\hat{t},\hat{\nabla}, \hat{B}_0, \hat{n}, \hat{T}\right)$ and $\left(\frac{-ic\delta\phi}{B_a R_0 V_{Ap}}, \frac{\delta A_{||}}{B_0R_0}, \frac{\delta n_e}{n_{ea}}\right)\to\left(\hat{V},\hat{Q},\hat{\delta n}_e\right)$, Eqs. \eqref{vor3}-\eqref{dne3} can be rewritten as following dimensionless form
	\begin{flalign}\label{vor}
		\begin{split}
			\hat{\omega}\hat{\nabla}\cdot\left(\frac{1}{\hat{V}_A^2}\hat{\nabla}_\perp \hat{V}\right)
			= \underbrace{\hat{\omega}_{*p,i}\hat\nabla\cdot\left(\frac{1}{\hat{V}_A^2}\hat{\nabla}_\perp\hat{V}\right)}_{Ion\ dia-drift}
			-\mathbf{\hat{B}_0}\cdot\hat{\nabla}\left[\frac{1}{\hat{B}_0^2}\hat{\nabla}\cdot\left(\hat{B}_0^2\hat{\nabla}_\perp \hat{Q}\right)\right]
			+\left(\hat{\nabla} \hat{Q}\times\mathbf{\hat{B}_0}\right)\cdot\hat{\nabla}\left(\frac{\hat{J}_{||0}}{\hat{B}_0}\right)
		\end{split},
	\end{flalign}
	
	\begin{flalign}\label{ohm}
		\begin{split}
			\hat{\omega}\hat{Q}
			=\frac{1}{\hat{B}_0}\mathbf{b_0}\cdot\hat{\nabla}\hat{V}
			\underbrace{+i\frac{\hat{T}_{e0}}{\hat{n}_{e0}\hat{B}_0}\mathbf{b_0}\cdot\hat{\nabla}\hat{\delta n_e}\left(\hat{\rho}_s\sqrt{\frac{\beta_{ea}}{2}}\right)
				+\hat{\omega}_{*n,e} \hat{Q}}_{Electron\ dia-drift}
			\underbrace{+i\frac{1}{4\pi}\frac{c^2}{V_{Ap}^2}\hat{\eta}_{||}\hat{\nabla}_\perp^2 \hat{Q}}_{Resistivity}
		\end{split},
	\end{flalign}
	
	\begin{flalign}\label{dne}
		\begin{split}
			\hat{\omega} \hat{\delta n_e}
			= -i\frac{\hat{q}_e\hat{n}_{e0}}{\hat{T}_{e0}}\hat{\omega}_{*n,e}\hat{V}\frac{1}{\hat{\rho}_s\sqrt{\frac{\beta_{ea}}{2}}}
		\end{split},
	\end{flalign}
	where $R_0$, $B_{ea}$, $n_{ea}$, $T_{ea}$, $V_{Ap} = B_{a}/\sqrt{4\pi n_{ea}m_p}$ and $\beta_{ea} = 8\pi n_{ea}T_{ea}/B_{a}^2$ denote the on-axis values of major radius, equilibrium magnetic field, equilibrium electron density, equilibrium electron temperature, proton Alfv\'en speed and electron pressure ratio, respectively. $\hat{\rho}_s = c\sqrt{m_pT_{ea}}/(eB_{a}R_0)$ and $m_p$ denotes the proton mass. Eqs. \eqref{vor}-\eqref{dne} can be casted into a matrix equation and implemented as one of the candidate models in MAS multi-level physics framework \cite{Bao2023}
	\begin{flalign}\label{EVP}
		\begin{split}
			\mathbb{A}\mathbf{X}=\hat{\omega}\mathbb{B}\mathbf{X}
		\end{split},
	\end{flalign}
	where $\mathbb{A}$ and $\mathbb{B}$ are the operator matrices, and $\mathbf{X} = \left[\hat{V},\hat{Q},\hat{\delta n_e}\right]^{T}$ denotes the eigenvector. In particular, MAS constructs operator matrices in the discretized form and solves Eq. \eqref{EVP} based on the eigenvalue problem (EVP) approach using MATLAB, which is able to obtain all unstable solutions for single-$m$ DTM problem efficiently. Regarding to the linear stability analysis, EVP approach is superior than the initial value problem (IVP) approach, since it can give the eigenvalue distribution of multiple branch modes, which brings convenience for instability identification and classification.

\end{sloppypar}
\end{document}